\documentclass[iop]{emulateapj}

\usepackage{natbib,aas_macros,amsmath}
\usepackage{multirow,color}

\newcommand{\rqq}{\textquotedblright}
\newcommand{\lqq}{\textquotedblleft}

\newcommand{\rrangle}{\right\rangle}
\newcommand{\llangle}{\left\langle}
\def\lrangle#1{\mbox{$\llangle #1 \rrangle$}}

\begin{document}
\shorttitle{Faint SMGs Revealed by Deep ALMA Observations}
\shortauthors{Ono et al.}
\slugcomment{Accepted for publication in ApJ}

\title{%
Faint Submillimeter Galaxies Revealed by Multifield Deep ALMA Observations:\\
Number Counts, Spatial Clustering, and A Dark Submillimeter Line Emitter  
}

\author{%
Yoshiaki Ono\altaffilmark{1}, 
Masami Ouchi\altaffilmark{1,2}, 
Yasutaka Kurono\altaffilmark{3}, 
and 
Rieko Momose\altaffilmark{1} 
}

\email{ono@icrr.u-tokyo.ac.jp}

\altaffiltext{1}{%
Institute for Cosmic Ray Research, The University of Tokyo,
Kashiwa, Chiba 277-8582, Japan
}
\altaffiltext{2}{%
Kavli Institute for the Physics andMathematics of the Universe 
(Kavli IPMU), WPI, The University of Tokyo, 
Kashiwa, Chiba 277-8583, Japan
}
\altaffiltext{3}{%
Joint ALMA Observatory, 
Alonso de Cordova 3107, Vitacura, Santiago 763-0355, Chile
}

\begin{abstract}
We present the statistics of faint submillimeter/millimeter galaxies (SMGs)
and serendipitous detections of a submillimeter/millimeter line emitter (SLE)
with no multi-wavelength continuum counterpart 
revealed by the deep ALMA observations. 
We identify faint SMGs with flux densities of $0.1-1.0$ mJy 
in the deep Band 6 and Band 7 maps 
of $10$ independent fields that reduce cosmic variance effects.
The differential number counts at $1.2$ mm are found 
to increase with decreasing flux density down to $0.1$ mJy. 
Our number counts indicate that 
the faint ($0.1-1.0$ mJy, or SFR$_{\rm IR} \sim 30-300 M_\odot$ yr$^{-1}$) SMGs
contribute nearly a half of the extragalactic background light (EBL),
while the remaining half of the EBL is mostly contributed 
by very faint sources with flux densities of 
$<0.1$ mJy (SFR$_{\rm IR} \lesssim 30 M_\odot$ yr$^{-1}$). 
We conduct counts-in-cells analysis with
the multifield ALMA data for the faint SMGs, 
and obtain a coarse estimate of galaxy bias, $b_{\rm g} < 4$. 
The galaxy bias suggests that 
the dark halo masses of the faint SMGs are $\lesssim 7 \times 10^{12} M_\odot$,
which is 
smaller than those of bright ($>1$ mJy) SMGs,
but 
consistent 
with abundant high-$z$ star-forming populations 
such as sBzKs, LBGs, and LAEs.
Finally, we report the serendipitous detection of SLE--1 
with continuum counterparts neither in our 1.2 mm-band 
nor multi-wavelength images  
including ultra deep \textit{HST}/WFC3 and \textit{Spitzer} data.
The SLE has a significant line at $249.9$ GHz 
with a signal-to-noise ratio of $7.1$.  
If the SLE is not a spurious source made by unknown 
systematic noise of ALMA, the strong upper limits of 
our multi-wavelength data suggest that 
the SLE would be a faint galaxy at $z \gtrsim 6$.
\end{abstract}


\section{Introduction} \label{sec:introduction}

In the past decades, 
it has been found that 
the amount of the cosmic infrared (IR) background 
is comparable to 
that of the cosmic optical background 
\citep{puget1996,fixsen1998,hauser1998,hauser2001,dole2006}.
The large amount of energy in the IR  
indicates that 
a significant fraction of the star formation in the universe is hidden by dust. 
Probing far-infrared (FIR) sources 
is key to a full understanding of galaxy formation history, 
and can provide strong constraints on models of galaxy formation 
\citep[e.g.,][]{granato2004,baugh2005,fontanot2007,shimizu2012,hayward2013}.

Considerable progress has been made in 
charting the abundance of FIR sources 
\citep[see the recent review of][]{casey2014}
and 
shown that 
the extragalactic background light (EBL) 
at  submillimeter and millimeter wavelengths
is largely contributed by dusty star-forming galaxies, 
the so-called submillimeter galaxies \citep[SMGs;][]{lagache2005}.
With a $15$-m dish, 
the James Clerk Maxwell Telescope (JCMT)
blank-field $850\mu$m submillimeter surveys
with 
Submillimeter Common User Bolometer Array \citep[SCUBA;][]{holland1999} 
have resolved $\sim 20-30${\%} of the $850\mu$m EBL
into distinct, bright SMGs with $S_{850\mu{\rm m}} > 2$ mJy
\citep[e.g.,][]{barger1998,hughes1998,barger1999,eales1999,eales2000,scott2002,borys2003,wang2004,coppin2006}.
Similar results have been obtained at $870\mu$m 
with the Large APEX Bolometer Camera \citep[LABOCA;][]{siringo2009} 
on the $12$-m APEX telescope \citep{weiss2009}.  
At $1.1$ mm, 
about $6-10$ {\%} of the EBL has been resolved into individual sources 
by deep surveys with the AzTEC camera \citep{wilson2008}
on both the JCMT 
\citep[e.g.,][]{perera2008,austermann2009,austermann2010}
and
the $10$-m Atacama Submillimeter Telescope Experiment 
\citep[ASTE; e.g.,][]{aretxaga2011,scott2010,hatsukade2011,scott2012}.

The biggest challenge 
for constructing the number counts of SMGs from such observations 
is the coarse spatial resolutions of the single-dish telescopes. 
Poor resolutions impose a fundamental limitation, 
the confusion limit \citep{condon1974}, 
on our ability to directly detect faint SMGs  
due to confusion noises. 
For instance, 
blank-field SCUBA surveys cannot reach the sensitivities 
required to identify the faint population below $2$ mJy at $850\mu$m. 
However, 
since the fraction of the millimeter and submillimeter EBL 
above $2$ mJy is not large, 
the total EBL is 
likely dominated by the population below the limit. 
Observations of massive galaxy cluster fields 
push the detection limits of intrinsic flux density toward fainter ones 
thanks to gravitational lensing effects 
\citep[e.g.,][]{smail1997,smail2002,cowie2002,knudsen2008,johansson2011,chen_cc2013}, 
but 
the positional uncertainties of the SMGs cause large uncertainties
in the amplifications and the intrinsic fluxes \citep{chen_cc2011}.

Another issue which arises from the poor resolutions is 
source blending; 
it is possible that several faint SMGs within a beam
appear as a single brighter SMG. 
Source blending 
possibly changes the shape of the number counts, 
most critically by mimicking a population of bright SMGs.
Multiplicity in a single-dish beam is also
expected from evidence of strong clustering among SMGs
\citep[e.g.,][]{blain2004,scott2006,weiss2009,hickox2012}.
In fact, 
interferometric observations have shown that 
close pairs are common among SMGs 
and 
a significant fraction of bright SMGs  
found by single-dish observations 
are resolved into multiple sources 
\citep[e.g.,][]{ivison2007b,wang2011,smolcic2012,barger2012,hodge2013,karim2013}.
although this issue is still under debate \citep[e.g.,][]{hezaveh2013,chen_cc2013b,koprowski2014}. 
To construct more reliable number counts 
down to flux densities of $< 1$ mJy, 
we need to conduct deep surveys with high angular resolution.

The Atacama Large Millimeter/submillimeter Array (ALMA) 
enables us to explore faint ($0.1-1.0$ mJy) SMGs 
without effect of confusion limit 
thanks to its high sensitivity and high angular resolution. 
\cite{hatsukade2013}
have shown the potential of ALMA; 
they have obtained number counts of 
unlensed faint SMGs down to sub-mJy level 
using ALMA. 
However, 
their ALMA data were originally obtained for their $20$ targets selected 
in one blank field, the Subaru/\textit{XMM-Newton} Deep Survey (SXDS) field \citep{furusawa2008} 
and the total survey area is not large, 
which may induce uncertainties in their measurements.

The physical properties of faint SMGs 
and 
their relationships with other galaxy populations found at similar redshifts 
have not yet been investigated well. 
The IR luminosities of the faint SMGs 
with $1.2$ mm flux densities of $0.1-1.0$ mJy 
are estimated to be $L_{\rm IR} \sim (1.5-15) \times 10^{11} L_\odot$, 
 if we adopt a modified blackbody with typical values for SMGs, 
i.e., spectral index of $\beta_{\rm d} = 1.5$  
and  
dust temperature of $T_{\rm d} = 35$ K \citep[e.g.,][]{kovacs2006,coppin2008b},  
located at $z=2.5$ \citep[e.g.,][]{chapman2005,yun2012}.  
In this case, 
from the estimated IR luminosities, 
their obscured star-formation rates (SFRs) are calculated to be 
SFR$_{\rm IR} \sim 30-300 M_\odot$ yr$^{-1}$ \citep{kennicutt1998b}. 
Recently, 
\textit{Herschel} observations have revealed that 
typical UV-selected galaxies such as Lyman-break galaxies (LBGs) 
have a median IR luminosity 
of $L_{\rm IR} \simeq 2.2 \times 10^{11} L_\odot$ 
\citep[][see also \citealt{leek2012,davies2013}]{reddy2012}, 
which is comparable to that of the faint SMGs. 
From a stacking analysis of \textit{Herschel} and ALMA data,  
\cite{decarli2014} have found that 
$K$-selected galaxies 
including star-forming BzK galaxies (sBzKs) have 
IR luminosities of $L_{\rm IR} = (5-11) \times 10^{11} L_\odot$. 
These results suggest that 
some of the faint SMGs might be FIR counterparts 
of UV- and/or $K$-selected galaxies.

The spatial clustering of SMGs is an important observable, 
since its strength can be used to estimate 
an average mass of their hosting dark matter haloes. 
\cite{blain2004} have measured the clustering length of SMGs brighter than 
$5$ mJy at $850\mu$m, 
and found that the clustering length is significantly larger than 
those of optical/UV color-selected galaxies at similar redshifts, 
suggesting that 
SMGs are hosted by very massive dark haloes, 
with dark halo masses of $M_{\rm DH} \sim 10^{13} M_\odot$ 
\citep[see also,][]{webb2003,weiss2009,hickox2012}. 
Although several studies have investigated 
the clustering properties of SMGs, 
little attempt has been made for measuring those of faint SMGs 
with sub-mJy flux densities. 
This is because 
the previous large area surveys with the single-dish telescopes 
cannot detect faint SMGs due to the confusion limit.

In this paper, 
we make use of multifield deep ALMA data, 
i.e., our own data for two independent fields 
and 
archival data with relatively long integration times,  
taken with the ALMA Band 6 and Band 7. 
Each field corresponds to a single primary beam area.
We focus on serendipitously detected sources 
other than the targeted sources. 
The combination of the results of the deep ALMA surveys 
and those of a wide area survey in the literature 
yields robust estimates on the number counts of SMGs 
over a wide range of flux densities ($\simeq 0.1-5$ mJy), 
which 
is currently 
one of the most reliable estimates on the abundance of SMGs.\footnote{
It is expected that 
the number counts of faint SMGs will be improved in the near future 
by combining results from ongoing ALMA deep field observations. 
}   
In addition, 
from the field-to-field scatter in their number counts,  
we carry out a pathfinder study for estimating 
the clustering properties of the faint SMGs.

Finally, 
we report the serendipitous detection of a line emitter at $1.2$ mm
using ALMA Band 6 data originally obtained for 
detecting [{\sc Cii}] emission from an extremely luminous Ly$\alpha$ blob at $z=6.595$, Himiko \citep{ouchi2013}. 
It is motivated by a recent discovery of 
a bright millimeter emission line 
beyond their target, nearby merging galaxies VV114 \citep{tamura2014}. 
Their spectral energy distribution (SED) analysis has shown that 
the detected line is likely a redshifted $^{12}$CO emission line 
from an X-ray bright galaxy at $z=2.467$, 
demonstrating that 
deep interferometric observations with high angular resolution 
can fortuitously detect emission lines 
not only from their main targets \citep{swinbank2012} 
but also from sources other than the targets \citep[see also,][]{kanekar2013b}.

The outline of this paper is as follows. 
After describing the ALMA observations 
and data reduction in Section \ref{sec:data},  
we perform source extractions and carry out simulations 
to derive the number counts of SMGs 
in Section \ref{sec:data_analysis}.
In Section \ref{sec:number_counts}, 
after we construct the number counts,  
we compare them with the previous observational results  
and model predictions,  
and estimate the contributions from the resolved sources 
to the EBL at 1.2 mm. 
In the next section, 
we present the results of our counts-in-cells analysis for faint SMGs. 
In Section \ref{sec:serendipitous_lines}, 
we report detections of serendipitous submillimeter emission lines  
in our ALMA data. 
A summary is presented in Section \ref{sec:summary}.

Throughout this paper, we assume a flat universe with 
$\Omega_{\rm m} = 0.3$, 
$\Omega_\Lambda = 0.7$, 
$n_{\rm s} = 1$, 
$\sigma_8 = 0.8$, 
and $H_0 = 70$ km s$^{-1}$ Mpc$^{-1}$. 
We use magnitudes in the AB system \citep{oke1983}.
Following the method by \cite{hatsukade2013}, 
we scale 
the flux density of a source observed at a wavelength different from $1.2$ mm 
to the flux density at $1.2$ mm 
by using a modified blackbody with typical values for SMGs as noted above.  
For the data that we analyze in this paper, 
we adopt the flux density ratios summarized in Table \ref{tab:ratios_fnu}. 
For the other data, we use 
$S_{\rm 1.2mm} / S_{870\mu{\rm m}} = 0.43$,  
$S_{\rm 1.2mm} / S_{\rm 1.1mm} = 0.79$, 
and 
$S_{\rm 1.2mm} / S_{\rm 1.3mm} = 1.25$.

\begin{deluxetable*}{ccccccc} 
\tablecolumns{7} 
\tablewidth{0pt} 
\tablecaption{
Survey Fields \label{tab:ratios_fnu}}
\tablehead{
\colhead{Map}
    & \colhead{Target} 
    & \colhead{$\lambda_{\rm obs}$}    
    & \colhead{$\nu_{\rm obs}$} 
    & \colhead{$\sigma$} 
    & \colhead{$S_{1.2{\rm mm}} / S_{\rm obs}$}   
    & \colhead{References} \\
\colhead{ }
    & \colhead{  }    
    & \colhead{(mm)}
    & \colhead{(GHz)}
    & \colhead{(mJy beam$^{-1}$)}
    & \colhead{  }    
    & \colhead{  }    \\
\colhead{ }
    & \colhead{ }
    & \colhead{(1)}
    & \colhead{(2)}
    & \colhead{(3)}    
    & \colhead{(4)}    
    & \colhead{(5)}    
}
\startdata 
$1$  & Himiko &  $1.16$  &  $259$  & $0.017$ &  $0.90$  & (a) \\ 
$2$  & NB921-N-79144 &  $1.22$  &  $245$  & $0.051$ &  $1.05$  & (b) \\ 
$3$  & LESS J033229.4$-$275619 &  $1.21$  &  $247$  & $0.075$ &  $1.03$  &  (c) \\ 
$4$  & CFHQS J0210$-$0456 &  $1.20$  &  $249$  & $0.031$ &  $1.00$  &  (d) \\ 
$5$  & CFHQS J2329$-$0301 &  $1.20$  &  $250$  & $0.021$ &  $1.00$  &  (d) \\ 
$6$  & ULAS J131911.29$+$095051.4 &  $1.16$  &  $258$  & $0.072$ &  $0.91$  &  (e) \\ 
$7$  & SDSS J104433.04$-$012502.2 &  $1.04$  &  $288$  & $0.088$ &  $0.68$  &  (e) \\ 
$8$  & SDSS J012958.51$-$003539.7 &  $1.04$  &  $288$  & $0.052$ &  $0.68$  &  (e) \\ 
$9$  & SDSS J231038.88$+$185519.7 &  $1.14$  &  $263$  & $0.058$ &  $0.87$  &  (e) \\ 
$10$  & SDSS J205406.49$-$000514.8 &  $1.15$  &  $261$  & $0.031$ &  $0.89$  &  (e) 
  \enddata 
\tablecomments{
(1) Observed wavelength. 
(2) Observed frequency. 
(3) The $1\sigma$ noise measured in each map 
before primary beam correction.
(4) Ratio of the flux density at $1.2$ mm, $S_{1.2{\rm mm}}$, 
to the observed flux density, $S_{\rm obs}$, 
on the assumption of a modified blackbody 
with typical values for SMGs. 
(5) (a) \cite{ouchi2013}; 
(b) R. Momose et al. in preparation; 
(c) \cite{nagao2012};  
(d) \cite{willott2013b}; 
(e) \cite{wang2013}. 
}
\end{deluxetable*} 

\section{ALMA Data} \label{sec:data}

We analyze continuum maps at around $1$ mm 
with high sensitivities and high angular resolutions, 
obtained in ALMA cycle 0 and cycle 1 observations.
In this section, 
we introduce the ALMA Band 6 data taken by our programs, 
and the other deep ALMA Band 6/Band 7 data that we use.

\subsection{Our Data} \label{subsec:our_data}

We use the ALMA data originally obtained by \cite{ouchi2013}, 
who targeted an extremely luminous Ly$\alpha$ blob at $z = 6.595$, Himiko. 
Deep ALMA Band 6 observations were carried out 
in 2012 July 15, 18, 28, and 31 
with a $16$ 12-m antenna array 
under the extended configuration of $36-400$ m baseline.  
To detect the redshifted [{\sc Cii}]$158$ $\mu$m line 
and simultaneously the dust continuum emission, 
they adopted four spectral windows  
with a bandwidth of $1875$ MHz. 
The central frequency of the four spectral bands is 
$250.24$ GHz,  $252.11$ GHz, 
$265.90$ GHz, and $267.78$ GHz. 
They used 3c454.3 and J0423$-$013 for bandpass calibrators 
and 
J0217$+$017 for a phase calibrator. 
Neptune and Callisto were observed as a flux calibrator. 
The total on-source integration time was $3.17$ hours.

We also use newly obtained ALMA Band 6 data (PI: R. Momose)
taken for a spectroscopically confirmed Ly$\alpha$ emitter (LAE)
at $z=6.511$, NB921-N-79144 \citep{ouchi2010}. 
They carried out deep ALMA Band 6 observations 
in 2013 June 18 and 19 
with $23$ antennas.  
They used four spectral windows, 
one with a bandwidth of $1875$ MHz 
and three with $2000$ MHz, 
to detect the redshifted [{\sc Cii}] line and dust continuum. 
The central frequency of the four spectral bands is 
237.62 GHz, 240.42 GHz, 
255.42 GHz, and 253.05 GHz. 
J0238$+$166 and J2258$-$279 were observed as a flux calibrator. 
The bandpass and phase were calibrated with J0204$-$1701 and J0215$-$0222, respectively. 
The total on-source integration time was $1.22$ hours.

The data were reduced with 
the Common Astronomy Software Applications \citep[{\sc casa};][]{mcmullin2007} 
package in a standard manner. 
Hereafter, 
the maps reduced from the data taken by \cite{ouchi2013} and R. Momose et al. in preparation 
are referred to as Map 1 and Map 2, respectively. 
The final synthesized beam sizes of the maps are  
$\sim 0 \farcs 6 - 0 \farcs 8$.  
The $1 \sigma$ noise of Map 1 (Map 2)
is 
$17$ ($52$) $\mu$Jy beam$^{-1}$ 
and is almost constant in the map 
uncorrected for the primary beam attenuation. 
Further details of the ALMA observations and 
sensitivities 
are summarized in \cite{ouchi2013} 
and 
will be presented in R. Momose et al. in preparation.
In what follows, 
we use the ALMA continuum maps 
within the primary beam model.

\subsection{Archival Data} \label{subsec:other_data}

To increase the number of SMGs for deriving the number counts, 
we take advantage of 
archival ALMA data 
that have been already public on the ALMA science archive.\footnote{\texttt{https://almascience.nrao.edu/aq/}} 
We include the ALMA Band 6 and Band 7 data 
with relatively long integration time 
taken by 
\cite{nagao2012}, 
\cite{willott2013b}, 
and 
\cite{wang2013}.

Their observations were conducted for 
spectroscopically confirmed quasars at $z \sim 5-6$. 
The number of the archival maps is eight in total: 
one from \cite{nagao2012}, 
two from \cite{willott2013b}, 
and five from \cite{wang2013}. 
Their targets 
and the central wavelengths of the continuum observations 
are summarized in Table \ref{tab:ratios_fnu}. 
The final synthesized beam sizes of the maps are $\sim 0 \farcs 6 - 1 \farcs 5$. 
The $1 \sigma$ noises of the maps uncorrected for primary beam attenuations 
are $\simeq 21 - 88$ $\mu$Jy beam$^{-1}$. 
For each map, 
we use the area contained in the primary beam.

\section{Data Analysis} 
\label{sec:data_analysis}

\begin{figure}
\begin{center}
   \includegraphics[scale=0.7]{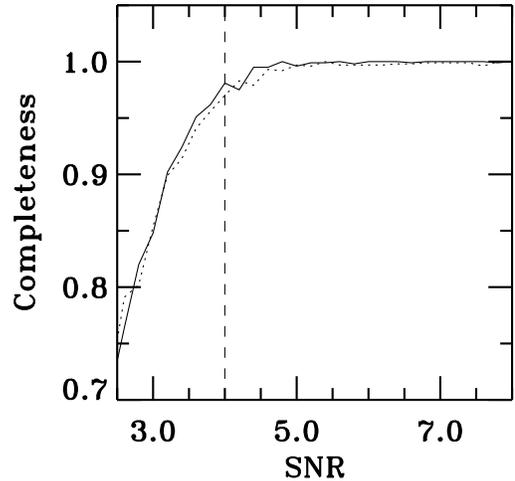}
 \caption[]
{
Completeness as a function of SNR 
estimated by Monte Carlo simulations. 
The solid curve and the dotted curve are  
the results of the simulations 
for Map 1 and Map 2, respectively. 
The vertical dashed line corresponds to 
the detection threshold we adopt. 
}
\label{fig:completeness}
\end{center}
\end{figure}

\begin{figure}
\begin{center}
   \includegraphics[scale=0.7]{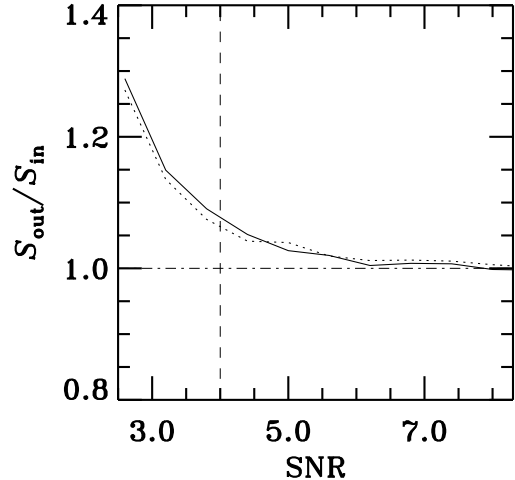}
 \caption[]
{
Flux boosting as a function of SNR 
estimated by Monte Carlo simulations.  
The solid curve and the dotted curve are  
the results of the simulations 
for Map1 and Map 2, respectively. 
The horizontal dot-dashed line corresponds to $S_{\rm out} = S_{\rm in}$.
The vertical dashed line shows 
the source detection threshold, SNR $=4$. 
}
\label{fig:fluxboost}
\end{center}
\end{figure}

\begin{figure}
\begin{center}
   \includegraphics[scale=0.7]{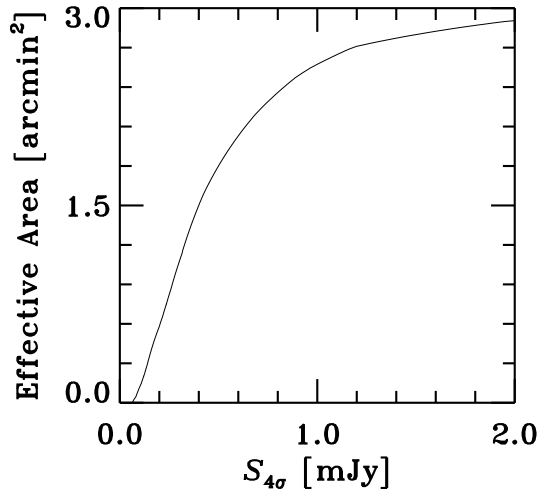}
 \caption[]
{
Total effective area of the ALMA maps analyzed in this study 
as a function of flux density,  
where a source with the flux density 
is detected at $> 4 \sigma$ 
in the primary beam corrected map. 
}
\label{fig:effective_area}
\end{center}
\end{figure}

In this section, 
we analyze the deep ALMA maps for the $10$ separate fields 
to derive the $1.2$ mm number counts of SMGs,  
basically following the method described in \cite{hatsukade2013}.

\subsection{Source Extraction} \label{subsec:alma_source_detection}

Source extractions are conducted on the ALMA maps 
before primary beam correction with 
SExtractor
version 2.5 \citep{bertin1996}. 
A collection of at least six contiguous pixels 
above the $1.8 \sigma$ noise level are identified as an object.  
We do not extract the targeted sources. 
Although this method can extract sources with relatively low SNRs, i.e., peak SNRs of $\geq 1.8$, 
we apply another threshold of peak SNRs to source catalogs to leave only reliable sources as we describe below.

Since 
the limit on significance levels for the source extractions is not high, 
some of our detections could be spurious caused by noise confusions. 
To estimate the fraction of spurious source detections,  
we use a negative ALMA map, 
i.e., a map multiplied by $-1$.  
The number of spurious sources is obtained 
by extracting sources from the negative map 
in the same manner as that for the positive map. 
We find that 
the angular distribution of the spurious sources is almost uniform. 
The number of spurious sources is less than 
that of positive sources at peak signal-to-noise ratio (SNR) $\gtrsim 3.5$, 
which is consistent with the results of \cite{hatsukade2013}.

We limit the catalogs to objects whose peak SNR is higher than $4.0$. 
In the $10$ ALMA maps, 
we detect 
eleven 
SMGs with SNRs of $4.1-6.1$. 
Their positions and SNRs are 
summarized in Table \ref{tab:catalog_faint_smgs}. 
Although two additional sources are detected with SNRs $> 4$ in Map 9,   
they are located 
where spurious sources can be made by 
the side-lobe effect of the bright quasar detected at the center of the map. 
Thus, we remove the two sources from our sample. 
All the SMGs 
appear to be 
point sources or at most marginally resolved.   
We find that 
two SMGs found in Maps 1 and 2 
have possible counterparts 
in the Subaru optical images \citep{furusawa2008}
and 
the \textit{Hubble Space Telescope} (\textit{HST}) 
near-infrared images \citep{ouchi2013}. 
Their detailed properties will be presented elsewhere. 
Note that 
an SMG detected in Map 5 has already been identified 
by \cite{willott2013b} as a blue galaxy at a moderate redshift 
(see their Figure 1).

\begin{deluxetable*}{ccccccc} 
\tablecolumns{7} 
\tablewidth{0pt} 
\tablecaption{Faint SMGs Identified in the $10$ ALMA Maps \label{tab:catalog_faint_smgs}}
\tablehead{
\colhead{Map}
    & \colhead{R.A.(J2000)}    
    & \colhead{Decl.(J2000)}    
    & \colhead{$S_{\rm obs}$}    
    & \colhead{SNR}    
    & \colhead{$S_{1.2{\rm mm}}$}    
    & \colhead{flag}    
\\
\colhead{ }
    & \colhead{ }    
    & \colhead{ }
    & \colhead{(mJy)}        
    & \colhead{ }
    & \colhead{(mJy)}        
    & \colhead{ }
\\
\colhead{ }
    & \colhead{ }
    & \colhead{ }
    & \colhead{(1)}
    & \colhead{(2)}    
    & \colhead{(3)}    
    & \colhead{(4)}
}
\startdata 
$1$ & 2:17:58.28 & $-$5:08:30.63 & $0.57 \pm 0.10$ & $5.7$ & $0.51 \pm 0.09$ & A \\ 
$2$ & 2:18:27.04 & $-$4:34:59.03 & $0.36 \pm 0.07$ & $4.9$ & $0.37 \pm 0.08$ & A \\ 
$2$ & 2:18:26.91 & $-$4:35:24.57 & $0.81 \pm 0.20$ & $4.2$ & $0.81 \pm 0.19$ & A \\ 
$3$ & 3:32:28.30 & $-$27:56:11.67 & $0.96 \pm 0.23$ & $4.1$ & $0.92 \pm 0.22$ & B \\ 
$4$ & 2:10:12.52 & $-$4:56:07.69 & $0.57 \pm 0.13$ & $4.3$ & $0.54 \pm 0.12$ & A \\ 
$4$ & 2:10:12.91 & $-$4:56:22.03 & $0.14 \pm 0.03$ & $4.1$ & $0.13 \pm 0.03$ & A \\ 
$5$ & 23:29:08.46 & $-$3:01:48.50 & $0.17 \pm 0.04$ & $4.9$ & $0.17 \pm 0.03$ & A \\
$5$ & 23:29:08.36 & $-$3:01:51.90 & $0.16 \pm 0.03$ & $6.1$ & $0.16 \pm 0.03$ & A \\
$6$ & 13:19:11.10 & 9:50:52.10 & $0.31 \pm 0.07$ & $4.3$ & $0.27 \pm 0.06$ & B \\ 
$9$ & 23:10:38.91 & 18:55:12.03 & $0.35 \pm 0.08$ & $4.5$ & $0.29 \pm 0.06$ & B \\
$9$ & 23:10:38.65 & 18:55:09.54 & $0.49 \pm 0.11$ & $4.6$ & $0.41 \pm 0.09$ & B  
  \enddata 
\tablecomments{
(1) Flux density at an observed frequency. 
(2) Signal-to-noise ratio of peak flux densities.  
(3) Estimated flux density at $1.2$ mm corrected for the effect of flux boosting. 
(4) A: no bright source with an SNR of $>10$ is detected in the map. 
B: although a bright quasar is detected at the center of the map, 
the position of the SMGs does not coincide with the positions of the side lobe. 
}
\end{deluxetable*} 

\begin{deluxetable}{cc} 
\tablecolumns{2} 
\tablewidth{0pt} 
\tablecaption{Differential Number Counts of SMGs at $1.2$mm \label{tab:number_counts}}
\tablehead{
\colhead{$S$}
    & \colhead{$\log (n)$}    \\
\colhead{(mJy)}
    & \colhead{([$\Delta \log S=1$]$^{-1}$deg$^{-2}$ )}    
}
\startdata 
\multicolumn{2}{c}{This Study} \\ \hline 
$0.13$  &  $4.83^{+0.37}_{-0.45}$ \\ 
$0.20$  &  $4.51^{+0.52}_{-0.76}$ \\ 
$0.32$  &  $4.42^{+0.30}_{-0.34}$ \\ 
$0.50$  &  $4.31^{+0.30}_{-0.34}$ \\ 
$0.79$  &  $3.75^{+0.37}_{-0.45}$ \\ 
\hline
\multicolumn{2}{c}{\cite{hodge2013} and \cite{karim2013}} \\ \hline 
$1.26$  &  $2.68^{+0.10}_{-0.11}$ \\ 
$2.00$  &  $2.97^{+0.07}_{-0.07}$ \\ 
$3.16$  &  $2.74^{+0.09}_{-0.09}$ \\ 
$5.01$  &  $1.40^{+0.52}_{-0.76}$ 
  \enddata 
\tablecomments{
The $1 \sigma$ uncertainties are calculated based on 
Poisson confidence limits \citep{gehrels1986}. 
}
\end{deluxetable} 

\subsection{Completeness and Flux Boosting} 

We calculate the detection completeness, 
which is the expected rate at which a source is detected in a map, 
to estimate the effects of noise fluctuations on source extractions. 
Since the source extractions are performed 
in the maps uncorrected for primary beam attenuation, 
the completeness estimations are 
conducted in the maps before primary beam corrections as well. 
We put a flux-scaled synthesized beam into a map 
as an artificial source.    
Since the noise level is almost constant in each map, 
the input position of an artificial source is randomly chosen in the map. 
We put artificial sources whose SNRs are in the range of $2 - 9$.
We perform source extractions in the same manner as 
that conducted for the actual maps (Section \ref{subsec:alma_source_detection}). 
From the fraction of recovered objects, 
we compute the completeness as a function of SNR 
(Figure \ref{fig:completeness}).

It has been reported that 
SNR limited source catalogs 
carry a selection bias from an overabundance of sources 
whose apparent fluxes are positively enhanced by noises \citep[e.g.,][]{hogg1998,scott2002}. 
From the simulations for estimating the detection completenesses, 
we also address this flux boosting issue. 
The results of our simulations are shown in Figure \ref{fig:fluxboost}, 
where we present the ratio of the extracted flux densities $S_{\rm out}$ to 
the input flux densities $S_{\rm in}$ as a function of SNR. 
The systematic differences 
between the output and input flux densities 
are less than only $10${\%} at SNR $> 4$.  
To obtain the de-boosted flux density, 
we divide the observed flux density of a detected source 
by the ratio $S_{\rm out} / S_{\rm in}$ 
at the SNR of the source estimated from the simulations.

\begin{figure*}
\begin{center}
   \includegraphics[scale=1.3]{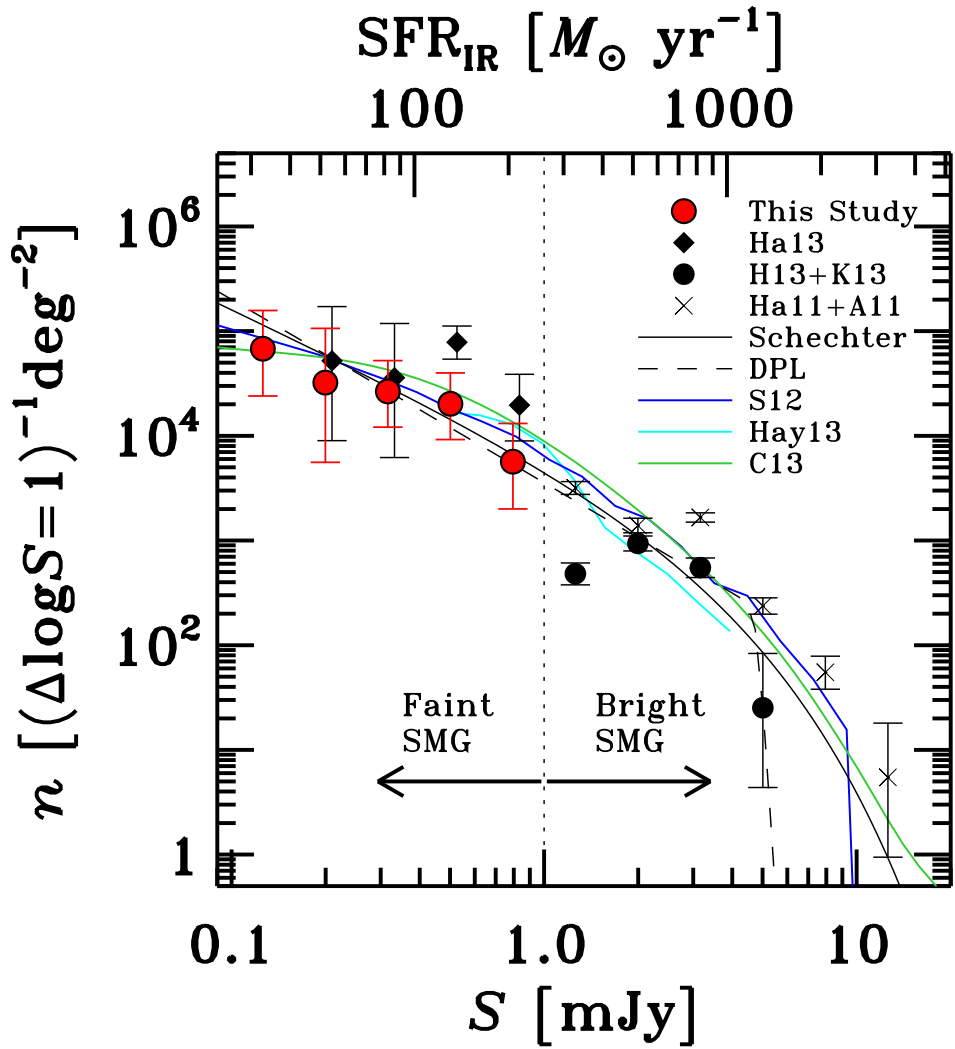}
 \caption[]
{
Differential number counts 
based on the results of ALMA observations. 
The number density of SMGs  
per $\Delta \log S = 1$ per unit square degree 
is plotted against flux density $S$ 
with corresponding SFR$_{\rm IR}$ on the upper $x$-axis. 
The red filled circles are 
the number counts derived from the faint SMGs found in this study. 
The black filled diamonds are 
the number counts estimated by using the source catalog of \cite{hatsukade2013}.  
Their flux densities have been shifted by $+0.05$ logarithmic units for clarity. 
The black filled circles are 
calculated based on the catalog of bright SMGs 
obtained from the ALMA Band 7 observations \citep{hodge2013,karim2013}. 
The crosses are 
the number counts of bright SMGs detected with ASTE AzTEC at 1.1 mm 
\citep{hatsukade2011,aretxaga2011}. 
The solid curve shows 
the best-fit Schechter function 
and 
the dashed curve is 
the best-fit DPL function. 
The blue curve shows 
the model predictions for the number counts 
based on cosmological hydrodynamic simulations 
with {\sc gadget}-3 \citep{shimizu2012}. 
The cyan curve corresponds to 
the model predictions calculated by combining 
a semi-empirical model with 3D hydrodynamical simulations 
and dust radiative transfer \citep{hayward2013}. 
The green curve is 
the theoretical predictions obtained by \cite{cai2013} 
based on their semi-analytical model. 
}
\label{fig:diff_n_count}
\end{center}
\end{figure*}

\begin{figure}
\begin{center}
   \includegraphics[scale=0.9]{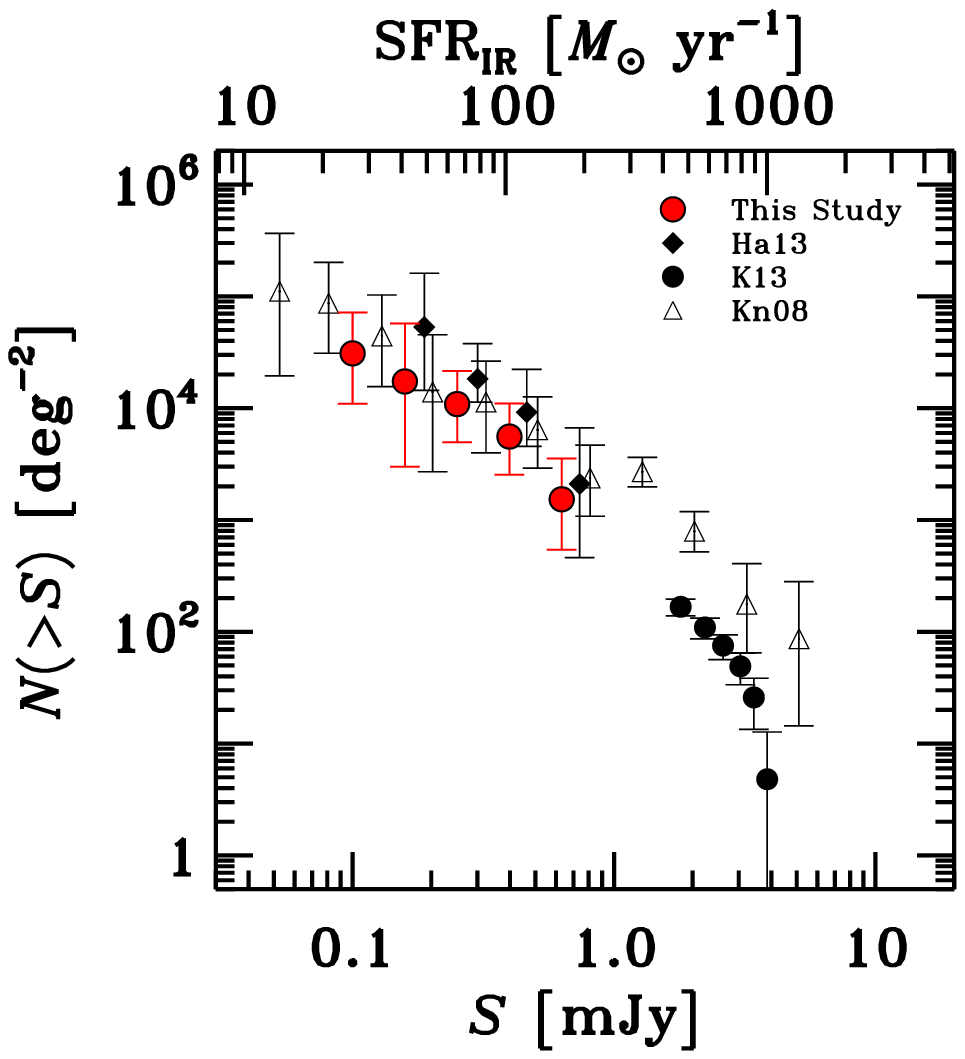}
 \caption[]
{
Cumulative number counts of SMGs. 
The number density of SMGs per unit square degree is plotted 
against flux density at $1.2$ mm, $S$, 
with corresponding SFR$_{\rm IR}$ on the upper $x$-axis. 
The red filled circles denote the cumulative number counts derived in this study.   
The black filled diamonds and circles 
are the cumulative number counts obtained by \cite{hatsukade2013} and \cite{karim2013}, respectively. 
The open triangles represent the number counts of SMGs reported by \cite{knudsen2008}.  
}
\label{fig:cumu_n_count}
\end{center}
\end{figure}

\section{Number Counts at 1.2 mm}
\label{sec:number_counts}

\subsection{Derivation of Differential Number Counts}  

By using the 
serendipitously discovered SMGs with SNRs of $> 4$, 
we derive the differential number counts of SMGs at $1.2$ mm. 
First, we estimate the effective area as a function of flux density 
(corrected for primary beam attenuation),  
since the primary beam response in a map depends on  
the distance from the center of the map.  
The derived effective area 
is shown in Figure \ref{fig:effective_area} 
as a function of $4 \sigma$ flux density, $S_{4 \sigma}$.

Using the results of our simulations described 
in Section \ref{sec:data_analysis}, 
we correct for the contamination of spurious sources and 
the effect of the incompleteness. 
A contribution from a detected source 
with an intrinsic flux density of $S$ 
to the number count, $\xi(S)$, is measured as 
\begin{equation}
\xi(S) 
	=  \dfrac{1 - f_{\rm c} (S)}{C(S) A_{\rm eff}(S)} 
\end{equation}
where 
$f_{\rm c}$ is the contamination fraction, 
$C$ is the completeness, 
and 
$A_{\rm eff}$ is the effective survey area. 
Then, we calculate a sum of the contributions, 
$n(S) = \sum \xi(S) / \Delta \log S$ 
in each logarithmic flux density bin 
$\log S \pm (1/2) \Delta \log S $, 
where $\Delta \log S = 0.2$, 
to obtain a logarithmic, differential number count.  
The obtained differential number counts are scaled to $\Delta \log S = 1$. 
To calculate the $1\sigma$ uncertainties, 
we take account of Poisson confidence limits \citep{gehrels1986} 
on the number of the SMGs in each flux density bin. 
In the calculations of the $1\sigma$ uncertainties, 
the average $\xi(S)$ factors cancel out.  
The derived number counts are summarized in Table \ref{tab:number_counts}. 
Note that the $1 \sigma$ uncertainties of the two flux density bins of 
$S$ (mJy) $=$ ($0.13$, $0.79$), ($0.20$, $5.01$), and ($0.32$, $0.50$) 
in the logarithmic scale are the same, 
since the numbers of the SMGs in the flux density bins are the same.

\subsection{Comparison with Previous Studies}
\label{subsec:results_number_counts}

The number counts of the faint 
($0.1-1$ mJy, or SFR$_{\rm IR} \sim 30-300 M_\odot$ yr$^{-1}$) SMGs 
revealed by the deep and high angular resolution ALMA observations 
are presented in Figure \ref{fig:diff_n_count}.  
The obscured SFR derived from the $1.2$ mm flux density 
by the same method as described in Section \ref{sec:introduction} 
is given on the upper $x$-axis. 
Since the total survey area of our data is not large, 
we find no source brighter than $1$ mJy. 
For the number counts of bright ($>1$ mJy) SMGs, 
we make use of the source catalog 
obtained from 
the recent follow-up observations of bright LABOCA sources 
with ALMA \citep{hodge2013,karim2013}. 
We use $91$ sources whose SNRs are higher than $4.0$ 
in Table 3 and Table 4 of \cite{hodge2013}. 
Note that 
the sample of the $91$ sources is not the same as 
that used by \cite{karim2013}, 
since  
we adopt the SNR threshold of $4.0$, 
which is the same as that applied for our sample 
but higher than that adopted by \cite{karim2013}. 
Considering 
the completeness and spurious detection rates 
estimated by \cite{karim2013}, 
we derive the number counts in the flux density range of $\simeq 1-5$ mJy, 
which are also presented in Table \ref{tab:number_counts}. 
We also construct the differential number counts 
using the source catalog of \cite{hatsukade2013} 
considering their estimates on 
the contamination rate of spurious sources, 
the incompleteness, 
and 
the effective survey area for each source.

From Figure \ref{fig:diff_n_count}, 
we find that 
the differential number counts increase 
with decreasing flux density 
down to $0.1$ mJy. 
We also find that 
the slope of the logarithmic number counts 
at sub-mJy flux densities is relatively small.  
The number counts increase 
by more than three orders of magnitude from $10$ mJy to $1$ mJy, 
while 
they increase 
by only about an order of magnitude from $1$ mJy to $0.1$ mJy. 
This indicates that 
the slope of the number counts of the faint SMGs is smaller 
than that of the bright SMGs.

Figure \ref{fig:diff_n_count} shows that  
the differential number counts of \cite{hatsukade2013} 
probe similarly faint flux densities, 
and are broadly consistent with our results 
within the $1 \sigma$ uncertainties. 
It should be noted that 
they used $20$ ALMA maps obtained in one blank field of SXDS  
while 
we compile the results of the $10$ independent fields. 
The effect of field-to-field variations 
on our measurements of number counts 
is expected to be smaller than that on the results of \cite{hatsukade2013}.

At bright flux densities around $4$ mJy, 
which corresponds to an obscured SFR of 
$\sim 1000$ $M_\odot$ yr$^{-1}$, 
we confirm that the differential number counts drop off, 
as \cite{karim2013} have already pointed out. 
The lack of large numbers of the bright SMGs  
implies that 
the obscured SFRs of dusty galaxies have a natural limit, 
which may be due to feedback processes  
by active galactic nuclei \citep[e.g.,][]{croton2006,bower2006}
and/or supernovae \citep[e.g.,][]{springel2003}, 
but also due to shortage of gas supply for star formation 
\citep{karim2013,hayward2013}. 
Note that 
the number counts at the bright end 
are derived only from the observations 
in the Extended Chandra Deep Field-South (ECDF-S), 
which could be affected by cosmic variance. 
This issue should be addressed by further observations 
of other large fields with ALMA.

Figure \ref{fig:diff_n_count} also presents
the results from single-dish AzTEC observations at $1.1$ mm. 
We confirm that 
the single-dish observational results  
significantly overestimate the number counts at the bright end 
likely due to the poor angular resolutions \citep[see also,][]{karim2013}.  
Observations with high angular resolution are required for 
avoiding bias due to source confusions.

\begin{deluxetable}{cccccc} 
\tablecolumns{6} 
\tablewidth{0pt} 
\tablecaption{Best-fit Parameters of Parametric Fits \\
to the Differential Number Counts of SMGs \label{tab:best_fit_params}}
\tablehead{
\colhead{Function}
    & \colhead{$S_\ast$}    
     & \colhead{$\phi_\ast$}   
     & \colhead{$\alpha$}   
     & \colhead{$\beta$}    
     & \colhead{$\chi^2_{\rm r}$$^{\dagger}$} \\
\colhead{ }
    & \colhead{(mJy)}    
    & \colhead{($10^2$ deg$^{-2}$)}
    & \colhead{ }    
    & \colhead{ }    
}
\startdata 
Schechter 
	& $2.3^{+2.3}_{-0.9}$  & $9.0^{+20.2}_{-7.3}$ & $-2.4^{+0.3}_{-0.3}$ & --- & $1.6$ \\
DPL 
	& $4.8^{+0.0}_{-0.8}$  & $1.0^{+0.5}_{-0.0}$ & $2.8^{+0.1}_{-0.2}$ & $46^{+40}_{-35}$ & $1.5$ 
  \enddata 
\tablecomments{
The best-fit parameters of Equation (\ref{eq:mod_Schechter_func})  
are estimated to be  
$S_0 \, ({\rm mJy}) = 2.3^{+2.2}_{-0.9}$, 
$N_0 \, ({\rm deg}^{-2}) = 380^{+1640}_{-340}$, 
and 
$\alpha' = -3.4^{+0.3}_{-0.3}$. 
Equation (\ref{eq:mod_Schechter_func}) 
is a Schechter functional form conventionally used in previous studies 
\citep[e.g,][]{coppin2006,knudsen2008,austermann2010}
but different from the original Schechter function. 
}
\tablenotetext{$\dagger$}{%
Reduced $\chi^2$. 
}
\end{deluxetable} 

\begin{figure*}
\begin{center}
   \includegraphics[scale=1.1]{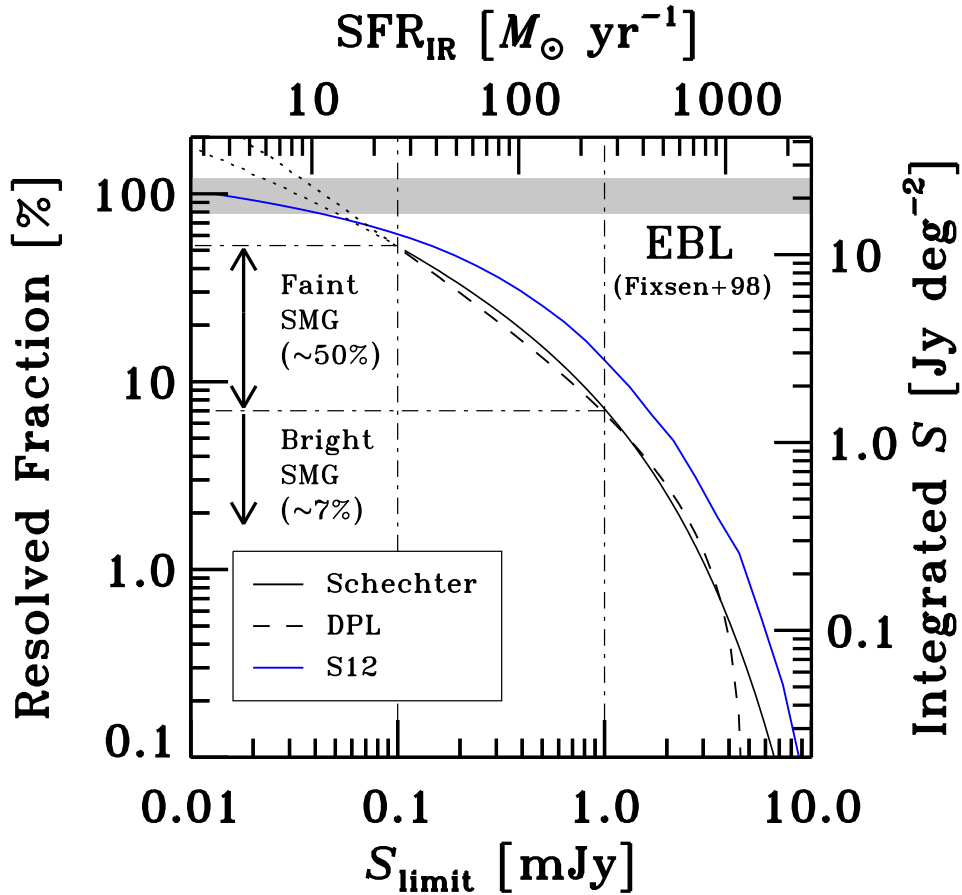}
 \caption[]
{
Fraction of resolved background light 
as a function of flux density limit at 1.2 mm, $S_{\rm limit}$ 
with corresponding SFR$_{\rm IR}$ on the upper axis.  
On the right axis, the absolute value of the integrated flux density, 
$\int^\infty_{S_{\rm limit}} S \phi (S) dS$, is shown. 
The black solid curve 
and 
the dashed curve 
correspond to the cases of our best-fit Schechter function  
and 
DPL function, 
respectively. 
The dotted curves represent  
simple extrapolations of the best-fit functions  
down to fainter $S_{\rm limit}$ than our survey limit.  
The blue curve is calculated from 
the predicted number counts of \cite{shimizu2012}. 
The gray hatched region is the extragalactic background light 
measured by the \textit{COBE} satellite \citep{fixsen1998}. 
}
\label{fig:cumulative_EBL}
\end{center}
\end{figure*}

Next we compare the observed number counts to theoretical predictions. 
In Figure \ref{fig:diff_n_count}, 
the number counts predicted by \cite{shimizu2012} 
are shown. 
They have performed cosmological hydrodynamic simulations 
with {\sc gadget}-3, 
implementing a simple dust absorption model 
and modified blackbody emission for the IR SEDs of galaxies. 
In their calculations, 
they have considered simulated galaxies 
with dark halo masses of $M_{\rm DH} > 10^{10} M_\odot$. 
Figure \ref{fig:diff_n_count} indicates that 
their predictions agree well with our observational results 
down to $\simeq 0.1$ mJy, 
and roughly reproduce the results 
at brighter flux densities of $\simeq 1-3$ mJy taken from the literature.  
This implies that 
the number counts of the SMGs 
with flux densities larger than $0.1$ mJy 
can be well explained by 
the abundance of galaxies with dark halo masses of $> 10^{10} M_\odot$. 
In Figure \ref{fig:diff_n_count}, 
the model predictions provided by \cite{hayward2013} 
are presented as well. 
They have adopted a hybrid approach 
by combining a semi-empirical model with 3D hydrodynamical simulations 
and 3D dust radiative transfer calculations. 
Their model predictions 
for the number counts are also broadly consistent with 
our observational results at $\simeq 0.5-1$ mJy 
and 
those at brighter flux densities in the literature.    
Figure \ref{fig:diff_n_count} also presents the theoretical results predicted by \cite{cai2013},  
who have adopted the semi-analytical model developed by \cite{granato2004}. 
They have combined 
a {\lqq}physically forward model{\rqq} evolving spheroidal galaxies and the associated AGNs  
with a {\lqq}phenomenological backward model{\rqq} 
for late-type galaxies and for the later AGN evolution. 
Their results 
are in good agreement with our observational results at the faint flux densities of $0.1-1$ mJy, 
and broadly consistent with those at the brighter flux densities.  
Their model predictions suggest that 
high-redshift ($z>1$) star-forming spheroidal galaxies 
dominate the number counts at the flux densities of $1-10$ mJy, 
while both high-redshift spheroidal galaxies 
and low-redshift galaxies (starbursts and normal late-type galaxies)  
contribute similarly to those at the faint flux densities in the range of $0.1-1$ mJy 
(Z. Y. Cai et al. 2013, private communication).

\cite{knudsen2008} have derived the cumulative number counts 
based on a sample of $15$ gravitationally lensed SMGs with flux densities 
below the blank-field confusion limit, i.e., $2$ mJy at $850$ $\mu$m. 
Since they do not provide the differential number counts, 
we derive the cumulative number counts to compare our results with theirs. 
Figure \ref{fig:cumu_n_count} 
shows the cumulative number counts derived from the sample of 
our faint SMGs and the bright SMGs taken from the literature \citep{hodge2013,karim2013}. 
We adopt a flux density step of $\Delta \log S = 0.2$. 
We calculate an $1 \sigma$ uncertainty in each bin 
based on Poisson confidence limits \citep{gehrels1986} 
of the differential numbers of the SMGs, 
following previous studies \citep[e.g.,][]{knudsen2008,hatsukade2013}. 
Figure \ref{fig:cumu_n_count} 
compares the cumulative number counts derived in this study 
with those obtained in the literature \citep{knudsen2008,karim2013,hatsukade2013}, 
which are taken from Table 4 of \cite{knudsen2008}, 
Table 1 of \cite{karim2013},    
and 
Figure 4 of \cite{hatsukade2013}. 
We find that 
our results are consistent with those of \cite{knudsen2008}, 
although their study is based on the sample of the lensed SMGs 
and thus have uncertainties in the intrinsic flux density estimates 
due to the positional uncertainties of the SMGs \citep{chen_cc2011}.  
This suggests that 
observations of massive galaxy cluster fields with the single-dish telescopes 
are also effective to investigate the abundance of faint SMGs, 
if source blending effect is not significant \citep{chen_cc2014}.

\subsection{Contributions to the Extragalactic Background Light}
\label{subsec:EBL}

We have constructed the $1.2$ mm number counts of SMGs 
based on the results of the multifield deep ALMA observations, 
which reduce biases due to field-to-field variations and source confusions. 
Using the improved number counts,  
we refine the estimates on the integrated flux density from resolved sources 
and their contributions to the EBL.

The integrated flux density 
from sources resolved into discrete objects 
can be calculated by 
integrating the product of the flux density and the number counts 
down to a flux density limit. 
In advance of integrating contributions from the resolved sources, 
we characterize the differential number counts 
using the following two functions often used in the literature.

One form is 
a Schechter function \citep{schechter1976}, 
\begin{equation}
\phi(S) dS
	= \phi_\ast \left( \dfrac{S}{S_\ast} \right)^\alpha 
		\exp \left( - \dfrac{S}{S_\ast} \right) 
		d \left( \dfrac{S}{S_\ast} \right),  
\label{eq:Schechter_func}
\end{equation}
where 
$\phi_\ast$, $S_\ast$, and $\alpha$ 
are the normalization, 
the characteristic flux density, 
and 
the faint-end slope, respectively.  
This functional form is motivated by 
the relationship between millimeter flux density 
and rest-frame FIR luminosity,  
which is nearly constant at high redshifts 
\citep{blain2002,coppin2006}. 
We define the logarithmic Schechter function $n (S)$ 
as $n(S) d (\log S) = \phi(S) dS$, i.e., 
\begin{equation}
\begin{split}
n(S)
	&= (\ln 10) \, \phi_\ast \,
		10^{(\alpha+1) (\log S - \log S_\ast)} \\
	& \hspace{5em}
		\times \exp \left( - 10^{(\log S - \log S_\ast)} \right), 
\end{split}
\end{equation}
and fit it to the differential number counts 
derived from the results of our observations 
as well as the previous observations.
In the fitting,
we take into account the following results: 
the number counts of the faint ($0.1-1.0$ mJy) SMGs 
presented in this paper, 
the number counts of the faint SMGs at $\simeq 0.2 - 1.0$ mJy 
constructed from the source catalog of \cite{hatsukade2013}, 
and 
the number counts of the bright ($>1$ mJy) SMGs 
derived from the source catalog of \cite{hodge2013} \citep[see also,][]{karim2013}. 
Note that 
we do not use the data point at $\simeq 1.2$ mJy, since it seems to be incomplete.  
Varying the three parameters, 
we search for the best-fitting set of ($\phi_\ast$, $S_\ast$, $\alpha$) 
that minimizes $\chi^2$. 
The best-fit parameters are 
shown in Table \ref{tab:best_fit_params}, 
and 
the best-fit Schechter function is plotted in Figure \ref{fig:diff_n_count}.

The original Schechter function described above 
is different from 
a Schechter functional form conventionally used in previous studies 
of submillimeter observations 
\citep[e.g.,][]{coppin2006,knudsen2008,austermann2010}, 
\begin{equation}
\phi(S) 
	= \dfrac{N_0}{S_0} S \left( \dfrac{S}{S_0} \right)^{\alpha'} 
		\exp \left( - \dfrac{S}{S_0} \right), 
\label{eq:mod_Schechter_func}
\end{equation}
where 
$N_0$ is the normalization, 
$S_0$ is the characteristic flux density, 
and 
$\alpha'$ is the faint-end slope. 
For the purpose of comparison, 
we derive the best-fit parameters of Equation (\ref{eq:mod_Schechter_func}) 
by a $\chi^2$ minimization fit 
to the observed differential number counts obtained in this study. 
The best-fit parameters are presented in the footnote of Table \ref{tab:best_fit_params}. 
\cite{knudsen2008} have derived the best-fit Schechter function 
of Equation (\ref{eq:mod_Schechter_func}) 
to the observed number counts of SMGs at $850$ $\mu$m 
based on their faint gravitationally lensed SMGs  
as well as bright SMGs from the SHADES survey \citep{coppin2006}. 
Our results 
are in agreement with those of \cite{knudsen2008}, 
except for the parameter of $\alpha'$. 
This is probably due to the difference in the number counts of the bright SMGs. 
For the number counts of the bright SMGs, 
we have used the results of the ALMA observations \citep{hodge2013,karim2013}. 
In contrast, 
\cite{knudsen2008} have used the results of the SHADES survey, i.e., 
the results from the single-dish observations, 
which cause an overestimate of the number counts at the bright flux densities 
due to the source blending issue, 
and 
make the slope of the number counts flatter apparently,  
as can be seen in Figure \ref{fig:cumu_n_count}.

The other form is 
a double power law (DPL) function 
\citep[e.g.,][]{scott2002,coppin2006}:  
\begin{equation}
\phi(S)
	= \dfrac{\phi_\ast}{S_\ast}
		\left[ \left( \dfrac{S}{S_\ast} \right)^\alpha + \left( \dfrac{S}{S_\ast} \right)^\beta \right]^{-1}, 
\label{eq:DPL_func}
\end{equation}
where 
the definition of $\phi_\ast$, $S_\ast$, and $\alpha$ are 
the same as those in equation (\ref{eq:Schechter_func}), 
and 
$\beta$ is the bright-end slope. 
The logarithmic DPL function $n (S)$ 
is defined as 
$n(S) d (\log S) = \phi(S) dS$, i.e., 
\begin{equation}
n (S)
	= \dfrac{(\ln 10) \phi_\ast }{ 10^{(\alpha-1) (\log S - \log S_\ast)} + 10^{(\beta-1) (\log S - \log S_\ast)} }. 
\end{equation}
Varying the four parameters, 
we search for the best-fitting DPL function that minimizes $\chi^2$. 
Table \ref{tab:best_fit_params} 
shows the best-fit set of the parameters. 
Figure \ref{fig:diff_n_count} shows the best-fit DPL function as well.

Using the best-fit parameter sets, 
we calculate integrated flux densities, 
$\int^\infty_{S_{\rm limit}} S \phi (S) dS$ 
down to a lower limit of flux density, $S_{\rm limit}$. 
Figure \ref{fig:cumulative_EBL} 
shows the integrated flux densities 
on the right axis, as a function of $S_{\rm limit}$.  
The integrated flux density down to the survey limit, 
$S_{\rm limit} = 0.1$ mJy, 
is calculated to be $\simeq 11$ Jy deg$^{-2}$, 
whichever of the two functional forms is adopted.  
Note that 
the EBL at 1.2 mm has been estimated  
to be $21.1^{+4.4}_{-4.6}$ Jy deg$^{-2}$ 
based on the observations by 
the Far Infrared Absolute Spectrophotometer (FIRAS) 
aboard the \textit{COBE} satellite \citep{fixsen1998}.

On the left axis of Figure \ref{fig:cumulative_EBL}, 
we show 
the fraction of the contributions from 
the resolved sources to the total EBL. 
We find that 
the faint ($0.1-1$ mJy, or SFR$_{\rm IR} \sim 30-300 M_\odot$ yr$^{-1}$) SMGs 
contribute nearly a half of the EBL, 
which is consistent with the results of \cite{knudsen2008}. 
Since the contributions from the bright SMGs are found to be less than $10${\%}, 
the remaining half of the EBL is mostly 
contributed by very faint SMGs with flux densities of $< 0.1$ mJy 
(SFR$_{\rm IR} \lesssim 30 M_\odot$ yr$^{-1}$). 
Although  
the resolved fraction of the EBL down to the survey limit 
is comparable to 
what \cite{hatsukade2013} have reported, 
the effects of field-to-field variations are reduced in our estimates. 
Note that 
the resolved fraction is significantly higher than 
those reported in the previous single-dish surveys 
at $1$ mm \citep[e.g.,][]{greve2004,hatsukade2011}.   
The higher sensitivities and higher resolutions of the ALMA maps 
enable us to push the survey limit down to the sub-mJy levels.

Interestingly, 
based on the theoretical prediction of \cite{shimizu2012},  
the integrated flux density calculated with the predicted number counts 
appears to converge to the EBL 
at $0.01$ mJy, which corresponds to SFR$_{\rm IR} \simeq$ $3$ $M_\odot$ yr$^{-1}$. 
This indicates that 
the EBL would be mostly explained 
by SMGs with $S = 0.01-10$ mJy.  
In this case, 
the contributions to the EBL 
from galaxies with obscured SFRs of $<3$ $M_\odot$ yr$^{-1}$ 
would be negligibly small. 
It is also inferred that, 
since the model predictions by \cite{shimizu2012}
have considered simulated galaxies with $M_{\rm DH} > 10^{10} M_\odot$, 
the dark halo masses of SMGs would be larger than $10^{10} M_\odot$ 
and  
the contributions to the EBL from 
small galaxies with $M_{\rm DH} < 10^{10} M_\odot$ would be negligible. 
This would be consistent with some theoretical studies 
that have suggested star formation quenching processes 
due to SNe feedback and/or UV radiation feedback 
work effectively in such small systems \citep[e.g.,][]{okamoto2010,hasegawa2013}.

\section{Counts-in-cells of Faint SMGs}
\label{sec:counts_in_cells}

The number of the detected SMGs differs among the maps, 
part of which could be induced by cosmic variance. 
From the field-to-field scatter in their number counts, 
we estimate the galaxy bias of the faint SMGs. 
With a sample of fields, 
the galaxy bias can be estimated from \citep{adelberger1998,robertson2010b}
\begin{equation}
b_{\rm g}^2
	\approx \dfrac{\sigma^2_N - \bar N}{\bar N^2 \sigma^2_V(z)}, 
\end{equation}
where 
$\bar N$ is the mean of source number counts per field, 
 $\sigma^2_N$ is the dispersion that includes 
contributions from both cosmic and Poisson variance, 
and 
$\sigma_V^2 (z)$ 
is the matter variance averaged over many survey volume $V$ 
\citep[e.g.,][]{mo2002}. 
To estimate an effective survey volume, 
we use a redshift distribution of SMGs  
obtained by follow-up spectroscopic observations \citep{chapman2005}. 
The redshift distribution has a long tail within $z = 1-4$, 
and its mean is $\lrangle{z} \simeq 2.5$.

\begin{figure}
\begin{center}
   \includegraphics[scale=0.9]{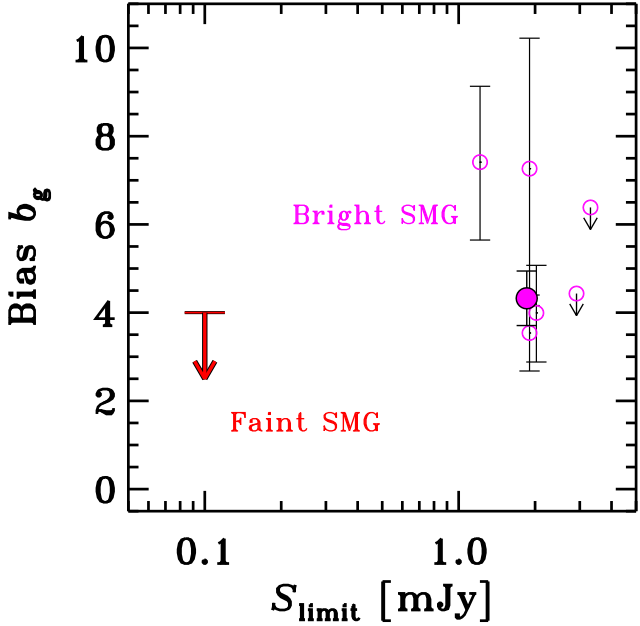}
 \caption[]
{
Bias of SMGs as a function of flux density limit, $S_{\rm limit}$. 
The red arrow 
represents the 
upper limit  
of faint SMGs analyzed in this paper. 
The magenta circles with error bars indicate bright SMGs 
\citep{webb2003,blain2004,weiss2009,hickox2012}, 
and the filled magenta circle corresponds to their average. 
The magenta circles with an arrow are the $1 \sigma$ upper limits 
of the bias of the bright SMGs estimated by \cite{williams2011}. 
}
\label{fig:bias_Snu}
\end{center}
\end{figure}

\begin{figure}
\begin{center}
   \includegraphics[scale=0.9]{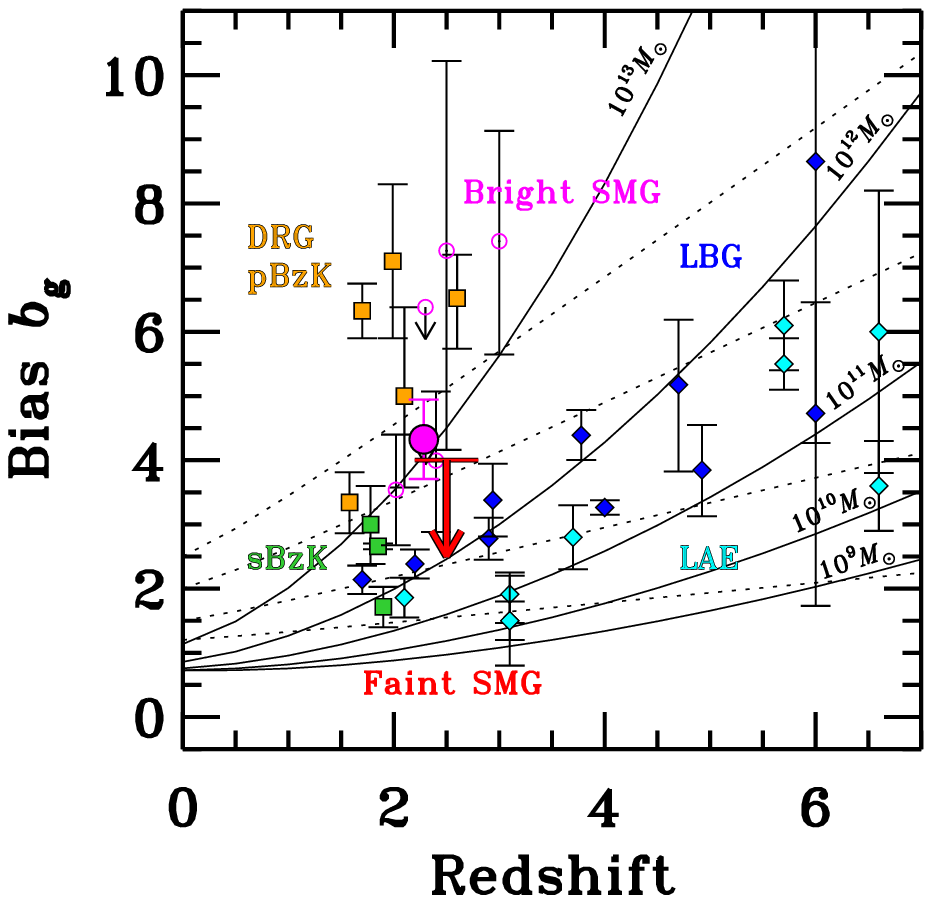}
 \caption[]
{
Bias of high redshift galaxies as a function of redshift. 
The red arrow 
represents the 
upper limit 
of faint SMGs analyzed in this paper. 
The open magenta circles with error bars indicate bright SMGs 
\citep{webb2003,blain2004,weiss2009,hickox2012}, 
and the filled magenta circle corresponds to their average. 
The open magenta circles with an arrow 
are the $1\sigma$ upper limits for the bright SMGs estimated by \cite{williams2011}. 
The blue diamonds represent 
LBGs and BX/BMs \citep{ouchi2004b,adelberger2005,leek2006,overzier2006}.
The cyan diamonds are LAEs \citep{gawiser2007,ouchi2010,guaita2009}.
The orange squares indicate 
DRGs and pBzKs \citep{grazian2006,quadri2007,blanc2008,furusawa2011,lin2012}. 
The green squares are 
sBzKs \citep{hayashi2007,blanc2008,furusawa2011}.
The solid curves are bias of dark haloes with a mass of 
$10^{13}$, $10^{12}$, $10^{11}$, $10^{10}$, and $10^{9} M_\odot$ from top to bottom, 
from \cite{sheth1999} 
in the case of one-to-one correspondence between galaxies and dark haloes. 
The dotted curves indicate the passive evolution of galaxy biases 
governed by their motion within the gravitational potential,  
assuming no merging \citep{fry1996}. 
}
\label{fig:bias_redshift}
\end{center}
\end{figure}

To calculate the mean and the dispersion of observed number counts per field, 
we need to set a flux density limit. 
This is because 
detection limits of flux density 
depend on positions in the maps 
due to primary beam attenuations. 
If we take a flux density limit of $0.25$ mJy 
with an uncertainty of a factor of two in flux density measurements, 
we obtain an $1\sigma$ upper limit of the galaxy bias of $b_{\rm g} < 4$. 
Note that 
if we take a flux density limit in the range of $\simeq 0.15 - 0.3$ mJy 
with an uncertainty of a factor of $2-3$, 
we obtain 
an $1\sigma$ upper limit of the galaxy bias in the range of $2-4$.

Figure \ref{fig:bias_Snu}
shows our estimate on the galaxy bias of the faint SMGs 
as well as the previous estimates for bright ($> 1$ mJy) SMGs 
\citep{webb2003,blain2004,weiss2009,williams2011,hickox2012}  
as a function of flux density limit.  
In the cases that 
galaxy biases are not presented and 
only the best-fit galaxy correlation functions are available in the literature, 
we calculate the galaxy bias from $b_{\rm g} = \sigma_{8,{\rm g}}/\sigma_8(z)$, 
where $\sigma_8(z)$ is a matter fluctuation in spheres of comoving radius of $8 h^{-1}$ Mpc, 
and $ \sigma_{8,{\rm g}}$ is a galaxy fluctuation,   
which is derived from \citep[e.g., eq.{[}7.72{]} in][]{peebles1993}, 
\begin{equation}
\sigma_{8,{\rm gal}}^2
	= \dfrac{72}{(3 - \gamma) (4 - \gamma) (6 - \gamma) 2^\gamma} 
		\left( \dfrac{r_0}{ 8 h^{-1} {\rm Mpc}} \right)^\gamma, 
\end{equation}
where 
$\gamma$ and $r_0$ are the parameters 
of the galaxy correlation function. 
In Figure \ref{fig:bias_Snu}, 
\cite{webb2003} and \cite{blain2004} 
have derived correlation lengths of 
$\simeq 11.5$ $h^{-1}$ Mpc and 
$\simeq 6.9$ $h^{-1}$ Mpc for the bright SMGs, 
which correspond to biases of $b_{\rm g} = 7.4$ and $4.0$, respectively. 
Similar results have been obtained for $870\mu$m LABOCA sources \citep{weiss2009,hickox2012}. 
The weighted-average bias
of the bright SMGs is calculated to be 
$b_{\rm g} = 4.3 \pm 0.6$. 
\cite{williams2011} have derived $1 \sigma$ upper limits of the correlation length 
based on their $1.1$ mm imaging, 
$r_0 \lesssim 6-8 h^{-1}$ Mpc ($11-12 h^{-1}$ Mpc) for bright SMGs down to $3.7$ ($4.2$) mJy.   
This correspond to an upper limit of the galaxy bias of $\lesssim 4.4$ ($6.4$), 
which is consistent with the weighted average. 
The galaxy bias of the faint SMGs estimated in this study 
appears to be lower than those of the bright SMGs reported in the literature.   
This difference indicates 
the clustering segregation with the FIR luminosity in SMGs. 
Similar clustering segregations 
with respect to the rest-frame UV/optical luminosities 
have been also found in local galaxies \citep[e.g.,][]{norberg2002} 
and 
high-redshift galaxies 
\citep[e.g.,][]{giavalisco2001,ouchi2004b,leek2006,hayashi2007,yoshida2008,leek2009,hildebrandt2009,bian2013,bethermin2014}.

It may be the case that 
some of the faint SMGs found in the archival data 
are physically related to the quasars at the center of the maps. 
However, 
since 
the redshift distribution of SMGs is substantially broad 
thanks to the negative $K$-correction \citep[e.g.,][]{chapman2005}, 
the effective survey volume is much larger than 
the volume observed around the quasar.  
In fact, 
a faint SMG that is detected in Map 5 
has already been identified 
at a much lower redshift than the target quasar \citep{willott2013b}, 
which indicates that 
at least the faint SMG has no physical relation to the quasar. 
In this study, 
we consider the faint SMGs detected in the archival maps 
have no relationship with the quasars, 
although we should keep in mind that 
the results might be biased by selecting the quasar fields. 
If this is the case, 
the galaxy bias of the faint SMGs would be smaller 
than our estimates.

Figure \ref{fig:bias_redshift} shows 
the galaxy biases of the faint and bright SMGs as a function of redshift,  
as well as the previous estimates for a variety of galaxy populations:  
$K$-selected galaxies 
\citep{grazian2006,hayashi2007,quadri2007,blanc2008,furusawa2011,lin2012} 
including passively-evolving BzK galaxies (pBzKs), 
sBzKs and distant red galaxies (DRGs), 
and 
UV-selected galaxies 
\citep{ouchi2004b,adelberger2005,overzier2006,leek2006,gawiser2007,ouchi2010,guaita2009} 
including BX/BM galaxies, LBGs and LAEs. 
At $z \sim 2.5$, 
the biases of DRGs and pBzKs 
appear to be higher than that of the faint SMGs, 
while 
$K$-selected galaxies with bluer UV colors (sBzKs) 
have galaxy bias values consistent with that of the faint SMGs.
The UV-selected galaxies at $z \sim 2-3$ 
have galaxy biases consistent with that of the faint SMGs 
as well. 
These results suggest that 
the dark halo masses of the faint SMGs 
might be 
comparable to 
those of sBzKs and UV-selected galaxies at $z \sim 2-3$. 
This implies that 
some of the faint SMGs might be their FIR counterparts, 
which is also suggested by the results of 
the recent \textit{Herschel} observations 
\citep{reddy2012,leek2012,decarli2014} 
as mentioned in Section \ref{sec:introduction}.

Dark halo masses of galaxies can be estimated with bias values 
in the frame work of the $\Lambda$CDM model. 
The solid curves in Figure \ref{fig:bias_redshift} 
indicate bias of dark haloes with a mass of 
$10^{13}$, $10^{12}$, $10^{11}$, $10^{10}$, and $10^{9} M_\odot$ from top to bottom, 
predicted by the \cite{sheth1999} model 
in the case of one-to-one correspondence between galaxies and dark haloes \citep[see also][]{mo2002}. 
Applying the model predictions, 
we estimate the dark halo mass of the faint SMGs 
to be roughly $\lesssim 7 \times 10^{12} M_\odot$. 
\cite{bethermin2013} have predicted that 
dark halos with $\sim 10^{12} M_\odot$ at $z \sim 2$ tend to host LIRG-like galaxies 
based on an abundance matching technique 
and their modeling approach 
that links stellar mass with star formation and infrared properties of galaxies (See their Figure 15), 
which is in agreement with our results.

To discuss possible present-day descendants of the faint SMGs, 
we show evolutionary tracks of dark haloes for galaxy-conserving models, 
which assume that the motion of galaxies is purely caused by gravity, 
and that galaxy merging does not occur \citep{fry1996}. 
In this case, 
the dark haloes of the faint SMGs 
would evolve into local galaxies with $b_g \lesssim 2$.  
This yields an upper limit of the galaxy bias evolution,  
since 
in a more realistic extended Press-Schechter formalism \citep[e.g.,][]{lacey1993}, 
evolutionary tracks are on average below those of the galaxy-conserving models \citep[e.g.,][]{ichikawa2007}. 
At the local Universe, 
the galaxy biases are 
in the range of $\sim 2-4$ for galaxy clusters \citep{bahcall2003}, 
about $2$ for galaxy groups \citep{girardi2000}, 
and 
in the range of $\simeq 1.1-2.2$ for the SDSS galaxy sample \citep{zehavi2005b,zehavi2011}. 
Our results imply that 
the faint SMGs could evolve into 
normal galaxies including Milky Way 
and might reside in galaxy groups at $z=0$.

Since the number of the pencil-beam survey fields is small, 
our estimates on galaxy biases have relatively large uncertainties. 
Moreover, 
the redshift distribution of the faint SMGs is unexplored 
and may differ from that of the bright SMGs. 
In fact, it has been reported that 
the redshift distributions of LIRGs and ULIRGs are different 
based on the sample of \textit{Spitzer} MIPS selected dusty sources. 
\citep[e.g.,][]{magnelli2011,murphy2011}. 
However, 
our coarse counts-in-cells analysis demonstrates 
the potential of ALMA for investigating the galaxy biases 
of faint SMGs, which have been poorly understood. 
Our results will be improved 
after a large number of deep ALMA maps 
and 
a redshift distribution of faint SMGs 
become available.

\begin{figure}
\begin{center}
   \includegraphics[scale=0.8]{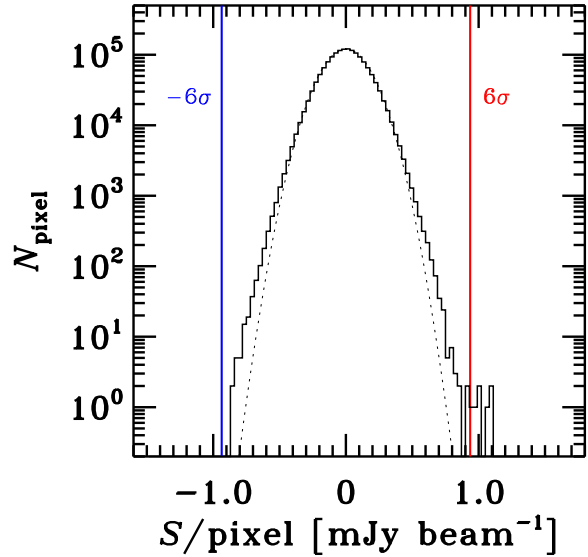}
 \caption[]
{
Histogram of flux density values (before primary beam correction) 
in the ALMA data cube with a frequency binning of $\Delta v = 100$ km s$^{-1}$. 
The black solid line represents the values of the signal map 
in the primary beam. 
The dotted curve shows the Gaussian function 
with rms of $\sigma \simeq 0.156$ mJy beam$^{-1}$ 
that is determined by a $\chi^2$ minimization fit to the histogram. 
The vertical blue and red lines correspond to $-6\sigma$ and $6 \sigma$, respectively. 
}
\label{fig:Dv100_fnu_histo}
\end{center}
\end{figure}

\begin{figure}
\begin{center}
   \includegraphics[scale=24.5]{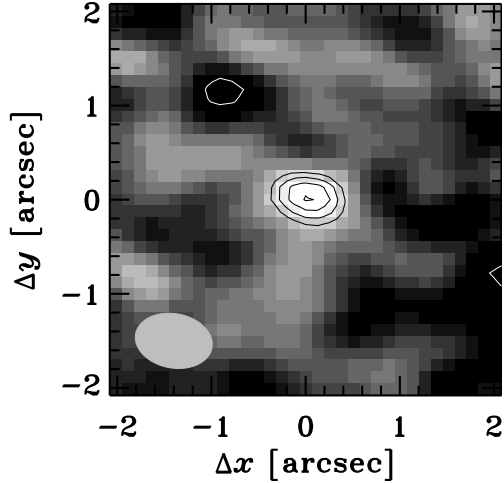}
 \caption[]
{
Emission-line map of a serendipitously detected line 
from a submillimeter line emitter, SLE--1,  
with 
a frequency binning of $\Delta v = 100$ km s$^{-1}$ 
and 
a spatial sampling of $0.13$ arcsec pixel$^{-1}$. 
The black contours corresponds to $4 \sigma$, $5 \sigma$, $6 \sigma$ and $7 \sigma$ levels, 
and the white contour denotes $-4 \sigma$ level. 
The shape of the synthesized beam is given at the bottom-left corner.  
}
\label{fig:band6_source4_line8}
\end{center}
\end{figure}

\begin{figure}
\begin{center}
   \includegraphics[scale=0.7]{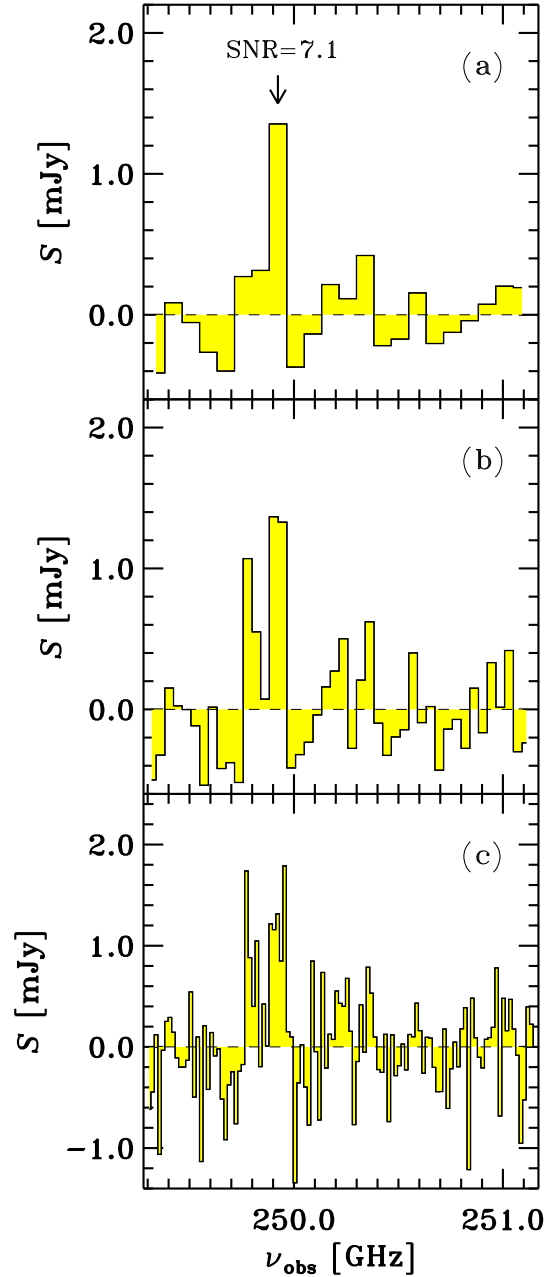}
 \caption[]
{
ALMA Band 6 spectra of SLE--1
extracted at the position of the peak emission  
with frequency binnings of  
(a) $\Delta v = 100$ km s$^{-1}$, 
(b) $\Delta v = 50$ km s$^{-1}$, 
and (c) $\Delta v = 20$ km s$^{-1}$.
The flux densities are corrected for the primary beam attenuation. 
The SNR of the peak flux density 
in the spectrum of $\Delta v = 100$ km s$^{-1}$ 
is $\simeq 7.1$.
}
\label{fig:spec1d_x258_y245}
\end{center}
\end{figure}

\section{Serendipitous Line Detection: 
Evidence for A Dark Submillimeter Line Emitter Population?} \label{sec:serendipitous_lines}

In the previous sections, 
we have analyzed the velocity-integrated continuum maps 
and 
presented the number counts and the spatial clustering 
of the serendipitously discovered faint SMGs. 
In this section, 
we conduct a blind search for line emitters at $\simeq 1.2$ mm 
in one of the data cube of Map 1 
taken with the ALMA Band 6.
This is a byproduct of our searching for [{\sc Cii}] emission from Himiko. 
Full results including the data cubes taken with the other three spectral windows will be presented in our future work. 
The central frequency and the bandwidth of the data are 
$250.24$ GHz and $1875$ MHz, respectively.

We create a data cube 
with $\Delta v =$ $100$ km s$^{-1}$ binning, 
to search for serendipitous emission lines within the primary beam. 
Figure \ref{fig:Dv100_fnu_histo} presents  
a histogram of flux density values 
uncorrected for the primary beam attenuation in the data cube. 
The histogram is well represented by a Gaussian distribution with 
an rms of 0.156 mJy beam$^{-1}$. 
From the histogram and the emission-line maps in the data cube, 
we find that 
one source is detected with an SNR of $>6$.
We also find that 
some negative sources show 
peak flux densities with SNRs in the range of $-6$ to $-5$, 
and 
no negative source has a peak flux density with an SNR of 
$< -6$. 
This indicates that 
the spurious detection rates for 
sources with SNRs in the range of $5-6$ are not negligible, 
while those for sources with SNRs 
$>6$
are substantially low.  
Note that 
the $1\sigma$ noise level in the frequency range of $\simeq 249.8 - 250.1$ GHz 
in which the source shows its peak flux density 
is higher than those outside of the range in the data cube. 
However, 
we confirm that 
this source shows a peak flux density 
which would not be explained by the noise.

\begin{deluxetable*}{ccccccccc} 
\tablecolumns{9} 
\tablewidth{0pt} 
\tablecaption{
Line Candidates for SLE--1  \label{tab:line_candidates}}
\tablehead{
\colhead{Species}
    & \colhead{Transition} 
    & \colhead{$\lambda_0$}    
    & \colhead{$\nu_0$}   
    & \colhead{$z$} 
    & \colhead{$L$} 
    & \colhead{$L'$} 
    & \colhead{$L_{\rm IR}^{(\rm exp)}$} 
    & \colhead{$S^{(\rm exp)}$} \\
\colhead{ }
    & \colhead{  }    
    & \colhead{($\mu$m)}
    & \colhead{(GHz)}    
    & \colhead{  }    
    & \colhead{(erg s$^{-1}$)}    
    & \colhead{($10^8$ K km s$^{-1}$ pc$^2$)}    
    & \colhead{($L_\odot$)}    
    & \colhead{($\mu$Jy)}    \\
\colhead{ }
    & \colhead{ }
    & \colhead{(1)}
    & \colhead{(2)}
    & \colhead{(3)}    
    & \colhead{(4)}    
    & \colhead{(5)}    
    & \colhead{(6)}    
    & \colhead{(7)}    
}
\startdata 
CO &  $J=3$ $\to$ $2$  &  $867$  &  $345.80$  &  $0.384$  &  $5.79 \times 10^{38}$ & $1.14$ & $1.9 \times 10^{10}$ & $32$ \\ 
CO &  $J=4$ $\to$ $3$  &  $650.3$  &  $461.04$  &  $0.845$  &  $3.90 \times 10^{39}$ & $3.24$ & $3.5 \times 10^{10}$ & $29$ \\ 
CO &  $J=5$ $\to$ $4$  &  $520.2$  &  $576.27$  &  $1.31$  &  $1.14 \times 10^{40}$ & $4.87$ & $3.8 \times 10^{10}$ & $27$ \\ 
CO &  $J=6$ $\to$ $5$  &  $433.6$  &  $691.47$  &  $1.77$  &  $2.41 \times 10^{40}$ & $5.94$ & $4.0 \times 10^{10}$ & $28$ \\ 
CO &  $J=7$ $\to$ $6$  &  $371.7$  &  $806.65$  &  $2.23$  &  $4.25 \times 10^{40}$ & $6.60$ & $5.2 \times 10^{10}$ & $37$ \\
CO &  $J=8$ $\to$ $7$  &  $325.2$  &  $921.80$  &  $2.67$  &  $6.72 \times 10^{40}$ & $6.98$ & $1.1 \times 10^{11}$ & $79$ \\
{[}{\sc Ci}{]} &  $^3P_2$ $\to$ $^3P_1$  &  $370.42$  &  $809.34$  &  $2.24$  &  $4.30 \times 10^{40}$ & $6.61$ & $3.0 \times 10^{12}$ & $2 \times 10^3$ \\ 
{[}{\sc Ci}{]} &  $^3P_1$ $\to$ $^3P_0$  &  $609.14$  &  $492.16$  &  $0.969$  &  $5.47 \times 10^{39}$ & $3.74$ & $1.4 \times 10^{11}$ & $1.1 \times 10^2$ \\ 
{[}{\sc Cii}{]}  &  $^3P_{3/2}$ $\to$ $^3P_{1/2}$  &  $157.74$  &  $1900.5$  &  $6.60$  &  $5.64 \times 10^{41}$ & $6.69$ & $4.7 \times 10^{10}$ & $44$ \\ 
{[}{\sc Oi}{]}  & $^3P_0$ $\to$ $^3P_1$ &  $145.53$  &  $2060.1$  & $7.24$ &  $6.98 \times 10^{41}$ &  $6.50$  & $6.5 \times 10^{12}$ & $6 \times 10^3$ \\ 
H$_2$O & $1_{11} \to 0_{00}$ & $269.27$ & $1113.34$ &  $3.46$  &  $1.23 \times 10^{41}$  &  $7.24$  &  $1.3 \times 10^{13}$  & $1 \times 10^{4}$ \\
H$_2$O & $2_{02} \to 1_{11}$ & $303.46$ & $987.93$ &  $2.95$  &  $8.42 \times 10^{40}$  &  $7.11$  &  $2.6 \times 10^{12}$  & $2 \times 10^3$ \\
H$_2$O & $2_{11} \to 2_{02}$ & $398.64$ & $752.03$ &  $2.01$  &  $3.31 \times 10^{40}$  &  $6.33$  &  $1.2 \times 10^{12}$  & $9 \times 10^2$ \\
H$_2$O & $2_{20} \to 2_{11}$ & $243.97$ & $1228.79$ &  $3.92$  &  $1.66 \times 10^{41}$  &  $7.27$  &  $4.9 \times 10^{12}$  & $4 \times 10^3$ \\
H$_2$O & $3_{12} \to 3_{03}$ & $273.19$ & $1097.37$ &  $3.39$  &  $1.17 \times 10^{41}$  &  $7.23$  &  $4.1 \times 10^{12}$  & $3 \times 10^3$ \\
H$_2$O & $3_{21} \to 3_{12}$ & $257.79$ & $1162.91$ &  $3.65$  &  $1.40 \times 10^{41}$  &  $7.26$  &  $3.6 \times 10^{12}$  & $4 \times 10^3$ \\
H$_2$O & $4_{22} \to 4_{13}$ & $248.25$ & $1207.64$ &  $3.83$  &  $1.57 \times 10^{41}$  &  $7.27$  &  $1.4 \times 10^{13}$  & $1 \times 10^{4}$ \\
H$_2$O & $5_{23} \to 5_{12}$ & $212.53$ & $1410.62$ &  $4.64$  &  $2.48 \times 10^{41}$  &  $7.19$  &  $7.4 \times 10^{12}$  & $7 \times 10^{3}$ \\
{[}{\sc Oiii}{]}  & $^3P_1$ $\to$ $^3P_0$ &  $88.36$  &  $3393.0$  & $12.6$ &  $2.45 \times 10^{42}$ &  $5.11$ & $>5.8  \times 10^{10}$ & $>35$ 
  \enddata 
\tablecomments{
(1) Rest-frame wavelength. 
For the species other than H$_2$O, we refer to Table 1 of \cite{carilli2013}. 
(2) Rest-frame frequency 
from Table 1 of \cite{carilli2013} and Table 1 of \cite{yang2013}. 
(3) Redshift. 
(4) Line luminosity. 
(5) Line luminosity calculated from eq.(\ref{eq:Lline_solomon}).  
(6) FIR luminosity estimated from the observed line flux. 
(7) Expected flux density at $259$ GHz ($\simeq 1.2$ mm) 
from the estimated FIR luminosity, 
assuming a modified blackbody with typical values for SMGs. 
}
\end{deluxetable*} 

Even if 
this source is 
not caused by statistical errors, 
it 
might be induced by unknown systematic noise effects 
in ALMA data cubes. 
If this is the case, 
we should be cautious in interpreting serendipitously detected 
single emission lines in ALMA data cubes. 
If 
it is 
not induced by systematic noise, 
the detected line is 
evidence for the existence of 
submillimeter/millimeter line emitters (SLEs). 
The SLE candidate is detected with an SNR of $7.1$.
If we use 
the $1\sigma$ noise level measured in the frequency bin where 
the SLE candidate is detected, the SNR of the peak flux density 
is estimated to be $5.3$.
Hereafter, 
the SLE candidate is referred to as SLE--1. 
Note that 
we also conduct blind line emitter searches  
in data cubes with smaller binnings, 
$\Delta v = 50$ and $20$ km s$^{-1}$,  
but we detect no source as reliable as SLE--1 with an SNR of $>7$.

Figure \ref{fig:band6_source4_line8} shows 
the spatial flux density distribution of SLE--1 at $\nu_{\rm obs} = 249.9$ GHz. 
SLE--1 has a point-source-like profile. 
The peak flux density corrected for the primary beam response is $\simeq 1.4$ mJy.

\begin{figure*}
\begin{center}
   \includegraphics[scale=10]{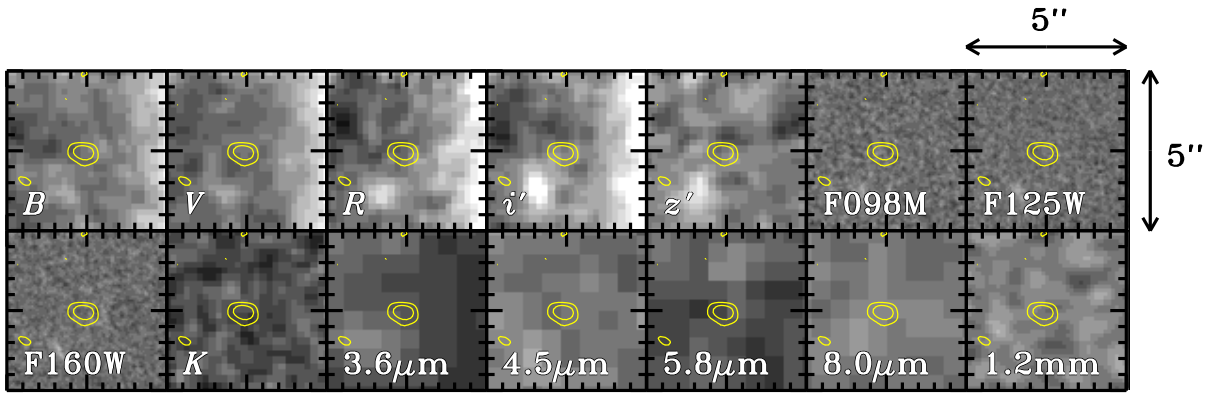}
 \caption[]
{
$5'' \times 5''$ multiwavelength images of SLE--1. 
Top panels show the images 
taken by the Subaru Suprime-Cam ($B$, $V$, $R$, $i'$, $z'$) 
and 
the \textit{HST} WFC3 (F098M, F125W). 
Bottom panels are the images 
taken by 
the \textit{HST} WFC3 (F160W), 
the UKIRT WFCAM ($K$), 
the \textit{Spitzer} IRAC ($3.6\mu$m, $4.5\mu$m, $5.8\mu$m, $8.0\mu$m) 
and 
the ALMA Band 6 ($1.2$ mm). 
Contours correspond to the $3\sigma$ and $5\sigma$ levels 
of the detected line at $249.9$ GHz. 
}
\label{fig:sme1_face}
\end{center}
\end{figure*}

\begin{figure*}
\begin{center}
   \includegraphics[scale=1.1]{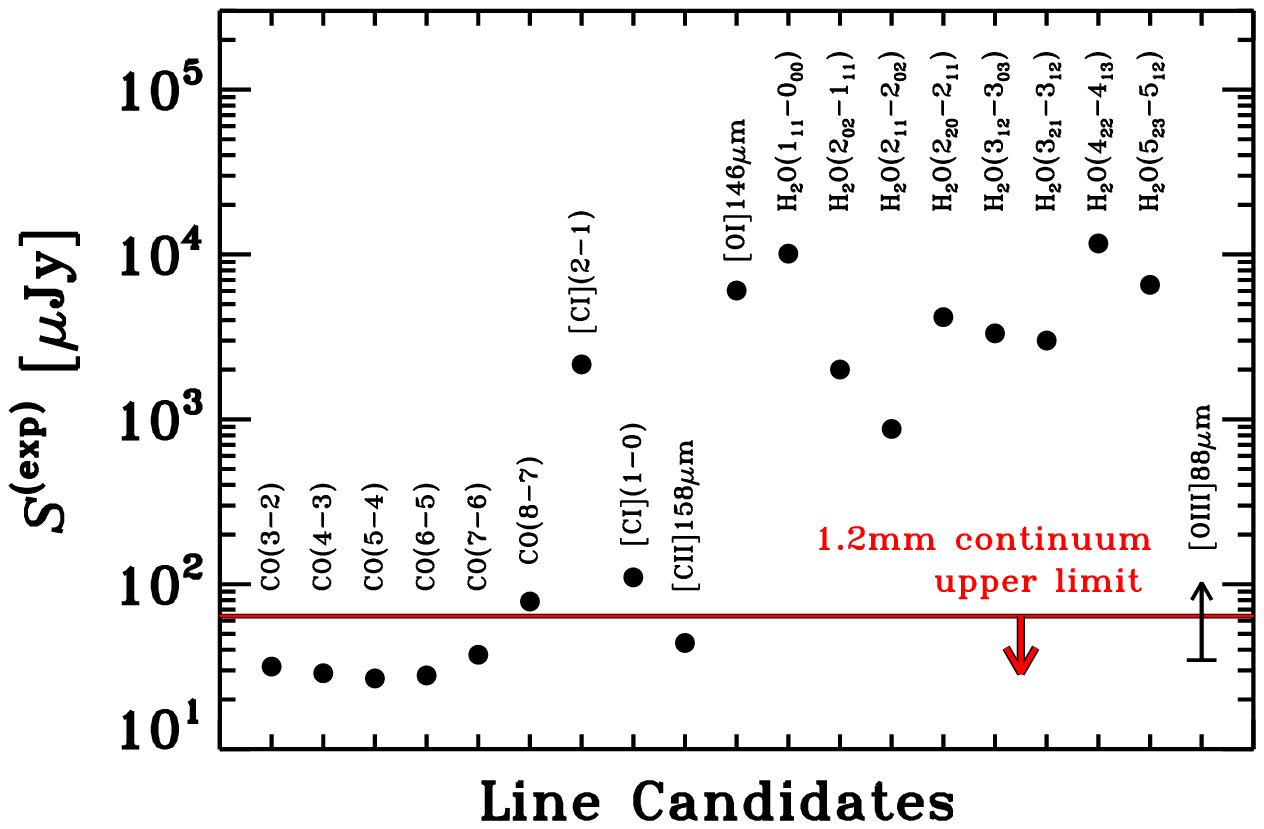}
 \caption[]
{
Expected $1.2$ mm continuum flux density of SLE--1
for the line candidates. 
The red horizontal line represents 
the $3\sigma$ upper limit of SLE--1 estimated from Map 1. 
}
\label{fig:Sexp_species}
\end{center}
\end{figure*}

In Figure \ref{fig:spec1d_x258_y245}, 
we plot spectra of SLE--1 
extracted at the position of the peak emission 
with three frequency binnings, 
$\Delta v =$ $100$, $50$, and $20$ km s$^{-1}$ from top to bottom.  
The spectra with $\Delta v =$ $50$, and $20$ km s$^{-1}$ 
appear to show a two-component profile at $\nu_{\rm obs} = 249.8-249.9$ GHz. 
Two-component line profiles have been reported for 
CO emission lines from 
star-forming galaxies at $z \sim 1.5-3$ \citep[e.g.,][]{coppin2007,daddi2010,tadaki2014b} 
and 
SMGs at $z \sim 1-5$ \citep[e.g.,][]{greve2005,weiss2005b,tacconi2006,tacconi2008,combes2012}. 
The detected line   
might have an analogous, two-component profile. 
However, 
in the spectrum with $\Delta v =$ $100$ km s$^{-1}$, 
one of the peaks at the lower frequency is smoothed out, 
which indicates that 
the SNR of the detected line is not high enough 
to conclude that the line has a two-component profile. 
Hereafter we consider the detected feature to be a single-peaked line.   
The flux of the detected line is calculated to be 
$(1.36 \pm 0.19) \times 10^{-1}$ Jy km s$^{-1}$, 
or $(1.13 \pm 0.16) \times 10^{-18}$ erg s$^{-1}$ cm$^{-2}$.

In our ALMA $1.2$ mm continuum map (Map 1), 
we detect no continuum emission 
at the position of SLE--1, 
as shown in the bottom-right panel of Figure \ref{fig:sme1_face}.  
The $3 \sigma$ upper limit on the $1.2$ mm continuum flux density of SLE--1  
is estimated to be $64$ $\mu$Jy.
Such a faintness suggests that SLE--1 would be a high-redshift source 
with an intrinsically strong FIR emission line.

\begin{figure}
\begin{center}
   \includegraphics[scale=9]{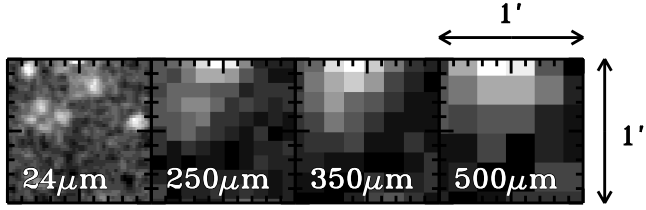}
 \caption[]
{
Mid- and far-infrared images of SLE--1 
taken by the \textit{Spitzer} MIPS ($24\mu$m) 
and 
the \textit{Herschel} SPIRE ($250\mu$m, $350\mu$m, $500\mu$m). 
SLE--1 is located at the center. 
Because 
the size of each panel is $1' \times 1'$, 
which is $12$ times larger than that in Figure \ref{fig:sme1_face}, 
no contour of the detected line is shown. 
}
\label{fig:sme1_face_FIR}
\end{center}
\end{figure}

Line candidates for SLE--1 are summarized in Table \ref{tab:line_candidates}. 
To examine the validity of the line candidates, 
first we roughly estimate the expected $1.2$ mm continuum flux densities of SLE--1 
from the observed line flux 
and test whether they are consistent 
with the upper limit estimated from Map 1. 
In the cases of the CO lines, 
we obtain 
CO(3--2) line fluxes in units of  Jy km s$^{-1}$ 
by adopting the CO excitation ladder found in M82 
\citep[][see also, \citealt{mao2000,ward2003,weiss2005} and Figure 4 of \citealt{carilli2013}]{weiss2005b}.\footnote{As noted later, 
we confirm that adopting the CO excitation ladder of Milky Way 
does not change our conclusions.} 
 Then, we derive CO(3--2) luminosities $L'_{{\rm CO}(3-2)}$  
in units of K km s$^{-1}$ pc$^2$ 
from the following equation 
\citep[][see also the reviews of \citealt{solomon2005} and \citealt{carilli2013}]{solomon1992}: 
\begin{equation}
L' \, ({\rm K} \, {\rm km} \, {\rm s}^{-1} {\rm pc}^2)
	= 3.25 \times 10^7 \, S_{\rm line} \Delta v \dfrac{D_L^2(z)}{(1+z)^3 \nu_{\rm obs}^2}, 
\label{eq:Lline_solomon}
\end{equation}
where 
$S_{\rm line} \Delta v$ is the observed flux of the line in units of Jy km s$^{-1}$, 
$D_L(z)$ is the luminosity distance in Mpc, 
$\nu_{\rm obs}$ in GHz is the observed frequency. 
Then we estimate the expected FIR luminosities, $L_{\rm IR}^{\rm (exp)}$, 
by using the relationship between 
the FIR luminosity and $L'_{{\rm CO}(3-2)}$ \citep{iono2009}, 
$\log L'_{{\rm CO}(3-2)} = \alpha \log L_{\rm IR} + \beta$,  
where ($\alpha$, $\beta$) $=$ ($0.93$, $-1.50$)
and 
$L'_{{\rm CO}(3-2)}$ and $L_{\rm IR}$ are 
in units of 
K km s$^{-1}$ pc$^2$ and $L_\odot$, respectively.\footnote{It should be noted that 
some recent studies of local LIRGs have shown that shocks can generate exceptionally high 
CO to continuum luminosity ratios \citep[][]{meijerink2013}, which we discuss later.  
} 
From the estimated FIR luminosities, 
we calculate the expected continuum flux densities at $\nu_{\rm obs} = 259$ GHz,  
which corresponds to $1.2$ mm, from \citep[e.g.,][]{ouchi1999}
\begin{equation}
S^{(\rm exp)}
	=  \dfrac{(1+z) L_{\rm IR}}{4 \pi D_L^2(z)} 
		\dfrac{\nu_0^{\beta_{\rm d}} B(\nu_0,T_{\rm d})}{\int \nu^{\beta_{\rm d}} B(\nu,T_{\rm d}) d\nu}, 
\label{eq:LIR}
\end{equation}
where $B(\nu,T)$ is the Planck function, 
and 
$\nu_0 = \nu_{\rm obs}(1+z)$.
We use 
$T_{\rm d} = 35$ K and $\beta_{\rm d} = 1.5$, 
which are typical values of SMGs as noted in Section \ref{sec:introduction}. 
Figure \ref{fig:Sexp_species} presents 
the expected $1.2$ mm continuum flux density for each line candidate. 
For the CO lines, 
the expected $1.2$ mm continuum flux densities are broadly consistent with 
the upper limit estimated in our ALMA continuum map. 
The higher CO transitions do not seem plausible, 
since the expected continuum flux densities become larger 
as the rotational quantum number $J$ increases.

If the detected line is [{\sc Ci}], 
the expected continuum flux densities are estimated to be 
$2$ mJy and $0.1$ mJy 
for [{\sc Ci}](2--1) and [{\sc Ci}](1--0), respectively, 
by adopting average ratios of 
$L_{{\rm [CI]}(1-0)} / L_{{\rm [CI]}(2-1)} = 2.7$ 
and 
$L_{{\rm [CI]}(1-0)} / L_{\rm IR} \sim 10^{-5}$ 
\citep{walter2011}. 
As shown in Figure \ref{fig:Sexp_species}, 
the estimated continuum flux density in the case of [{\sc Ci}](2--1)
is more than an order of magnitude brighter than 
the upper limit of Map 1, 
while 
the estimated value for [{\sc Ci}](1--0)
is consistent with the upper limit within a factor of two.

In the case that the detected line is [{\sc Cii}]$158\mu$m, 
we assume an average ratio of 
$L_{{\rm [CII]}} / L_{\rm IR} \sim 3.1 \times 10^{-3}$ 
derived by \cite{stacey2010b} for star-forming galaxies at $z = 1-2$. 
The expected $1.2$ mm continuum flux density is 
consistent with the upper limit estimated from the continuum map 
(Figure \ref{fig:Sexp_species}).

In the case of [{\sc Oi}]$146\mu$m, 
we adopt the relation between $L_{\rm IR}$ and the line luminosity, 
$\log L_{\rm IR} = \alpha + \beta \log L_{[{\rm OI}]145\mu{\rm m}}$, 
where ($\alpha$, $\beta$) $=$ ($1.75$, $1.34$)  
obtained by \cite{farrah2013}.  
In this case, 
the expected $1.2$ mm continuum flux density 
is two orders of magnitude brighter than the upper limit, 
suggesting that [{\sc Oi}]$146\mu$m is unlikely.

Recently, 
a series of detections of non-maser H$_2$O emission lines 
have been reported for high-redshift galaxies 
\citep[e.g.,][]{omont2011,vanderwerf2011,lupu2012,omont2013,riechers2013b}.  
We estimate the expected $1.2$ mm continuum flux densities 
in the cases that the detected line is H$_2$O 
by using the relations  
between the H$_2$O line luminosities and the FIR luminosity 
for starburst galaxies 
with $L_{\rm IR} \sim (1-300) \times 10^{10} L_\odot$ 
derived by \cite{yang2013}. 
As shown in Figure \ref{fig:Sexp_species}, 
the estimated $1.2$ mm continuum flux densities are far brighter than 
the upper limit of the ALMA continuum data, 
which suggests that 
the detected line is not H$_2$O.

\begin{figure}
\begin{center}
   \includegraphics[scale=27]{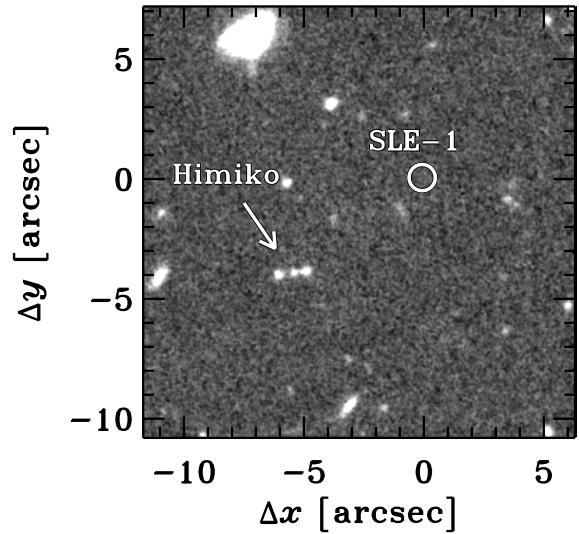}
 \caption[]
{
\textit{HST} WFC3 F125W image of SLE--1. 
The position of SLE--1 and Himiko clumps 
are marked with 
a circle 
and an arrow, respectively. 
To mark their positions, 
we use the CANDELS UDS astrometry, 
although \cite{ouchi2009} have used the astrometry of the SXDS version 1.0 catalog.
}
\label{fig:wfc3_f125w_sme}
\end{center}
\end{figure}

For the line candidates of SLE--1, 
we have investigated 
whether the expected $1.2$ mm continuum flux densities from the observed line flux 
are consistent with 
the upper limit obtained from Map 1. 
We have found that 
the estimated $1.2$ mm continuum flux densities are 
broadly consistent with the observed upper limit 
in the cases of 
the CO lines with the upper $J$ levels of $3-8$, 
[{\sc Ci}]($1$--$0$), 
and 
[{\sc Cii}]$158\mu$m. 
To further examine the validity of the lines,  
we search for possible counterparts in multiwavelength data. 
Figures \ref{fig:sme1_face} and \ref{fig:sme1_face_FIR} 
show SLE--1 in the multiwavelength data:  
the Subaru Suprime-Cam optical $BVRi'z'$ data \citep{furusawa2008}, 
the UKIRT WFCAM near-infrared $JHK$ data 
from the UKIDSS Ultra Deep Survey \citep[UDS;][]{lawrence2007,warren2007}, 
the \textit{HST} near-infrared data taken with 
the WFC3 F098M, F125W, and F160W filters \citep{ouchi2013} 
to which the data obtained by the Cosmic Assembly Near-IR Deep Extragalactic Legacy Survey 
\citep[CANDELS;][]{grogin2011,koekemoer2011} are added, 
the \textit{Spitzer} IRAC $3.6\mu$m and $4.5\mu$m data \citep[SEDS;][]{ashby2013} 
as well as the $5.8\mu$m and $8.0\mu$m, 
and MIPS $24\mu$m images (SpUDS; PI: J. Dunlop), 
and 
the \textit{Herschel} SPIRE $250\mu$m, $350\mu$m, and $500\mu$m data 
\citep[HerMES;][]{oliver2012}.
However, 
we find no detectable source at the position of SLE--1 
in the multiwavelength images. 
In addition, 
SLE--1 has no counterpart in 
the \textit{Galaxy Evolution Explorer} (\textit{GALEX}) FUV/NUV data \citep{nakajima2011}
and  
the \textit{XMM-Newton} data \citep{ueda2008}. 
The $3 \sigma$ upper limits on the flux densities of SLE--1 
at wavelengths from the optical to $1.2$ mm   
are summarized in Table \ref{tab:SED_upper_limits}. 
Figure \ref{fig:wfc3_f125w_sme} shows  
a cutout of the WFC3 F125W image 
with the location of SLE--1.

\begin{deluxetable*}{cccc} 
\tablecolumns{4} 
\tablewidth{0pt} 
\tablecaption{Upper Limits on the flux densities of SLE--1 \label{tab:SED_upper_limits}}
\tablehead{
\colhead{Instrument}
    & \colhead{Band/Wavelength}    
     & \colhead{Magnitude/Flux Density}   
    & \colhead{Remark}    \\
\colhead{ }
    & \colhead{ }    
    & \colhead{($3 \sigma$ limit, total)}
    & \colhead{ }
}
\startdata 
Suprime-Cam 	& $B$  &  $28.3$ mag  & (a) \\
Suprime-Cam 	& $V$  &  $28.0$ mag  & (a) \\
Suprime-Cam 	& $R$  &  $27.8$ mag  & (a) \\
Suprime-Cam 	& $i'$  &  $27.7$ mag  & (a) \\
Suprime-Cam 	& $z'$  &  $27.2$ mag  & (a) \\
WFC3 		& F098M  &  $28.4$ mag  & (a) \\
WFC3 		& F125W  &  $28.5$ mag  & (a) \\
WFCAM 		& $J$  &  $26.0$ mag  & (a) \\
WFC3 		& F160W  &  $28.1$ mag  & (a) \\
WFCAM 		& $H$  &  $25.5$ mag  & (a) \\
WFCAM		& $K$  &  $25.8$ mag  & (a) \\
IRAC 		& $3.6$ $\mu$m  &  $25.2$ mag  & (b) \\
IRAC 		& $4.5$ $\mu$m  &  $25.2$ mag  & (b) \\
IRAC 		& $5.8$ $\mu$m  &  $22.0$ mag  & (c) \\
IRAC 		& $8.0$ $\mu$m  &  $21.8$ mag  & (c) \\
MIPS 		& $24$ $\mu$m  &  $19.8$ mag  & (c) \\
SPIRE 		& $250$ $\mu$m  &  $6.7$ mJy  & (d) \\
SPIRE 		& $350$ $\mu$m  &  $5.6$ mJy  & (d) \\
SPIRE 		& $500$ $\mu$m  &  $8.0$ mJy  & (d) \\
ALMA 		& $1.2$ mm &  $64$ $\mu$Jy  & (e)
  \enddata 
\tablecomments{
(a) Measured within a $2 \times $ FWHM diameter aperture 
and corrected to total magnitude in a similar manner to \cite{mclure2012b}, 
assuming a point source and that the aperture depth is $0.3-0.4$ mag deeper. 
(b) Calculated over a $2 \farcs 4$ diameter aperture 
and corrected to total magnitude using the offset values listed in Table 2 of \cite{ashby2013}. 
(c) Measurements obtained in \cite{ouchi2009}. 
(d) Taken from Table 5 of \cite{oliver2012}. 
(e) Corrected for primary beam attenuation.
}
\end{deluxetable*} 

\begin{figure*}
\begin{center}
   \includegraphics[scale=0.7]{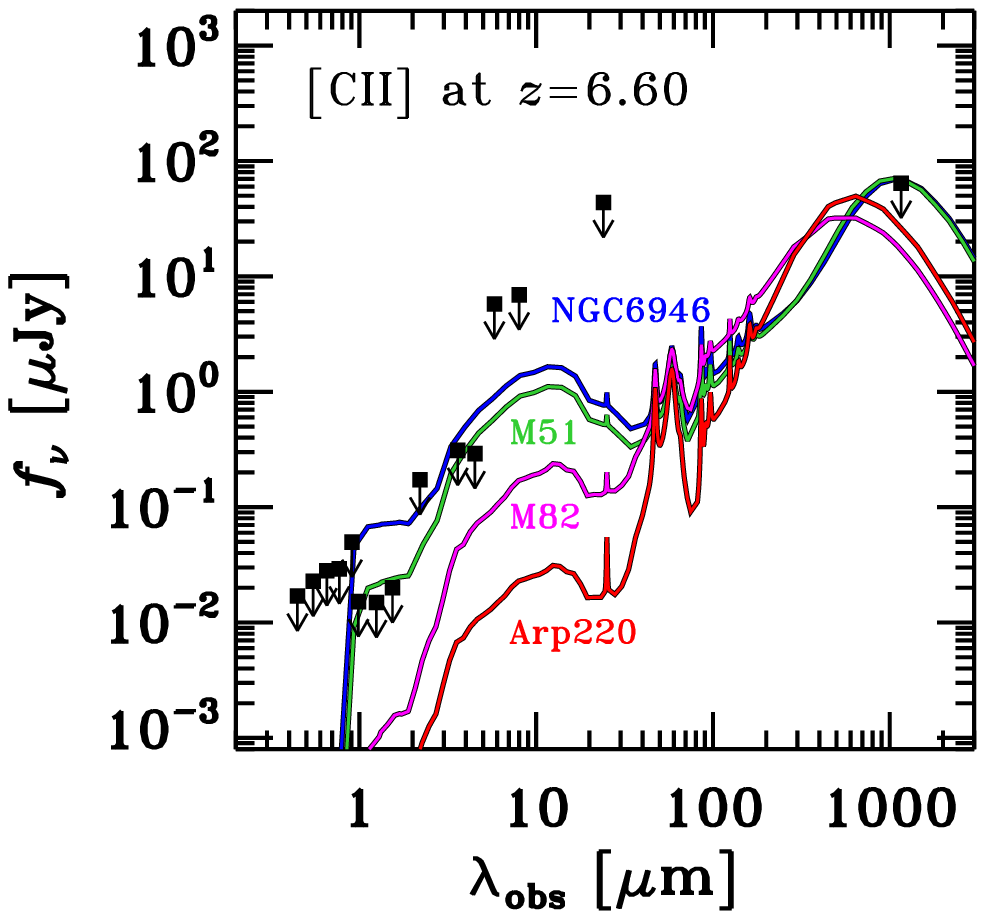}
   \includegraphics[scale=0.7]{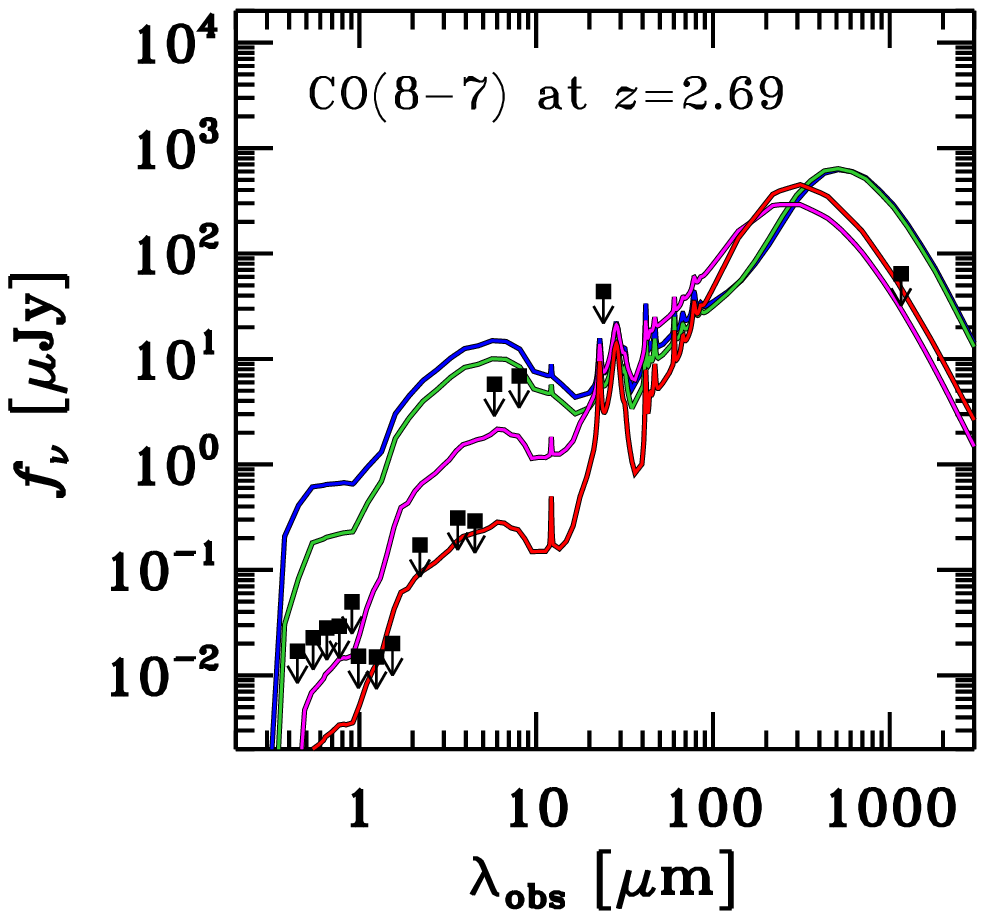}
   \includegraphics[scale=0.7]{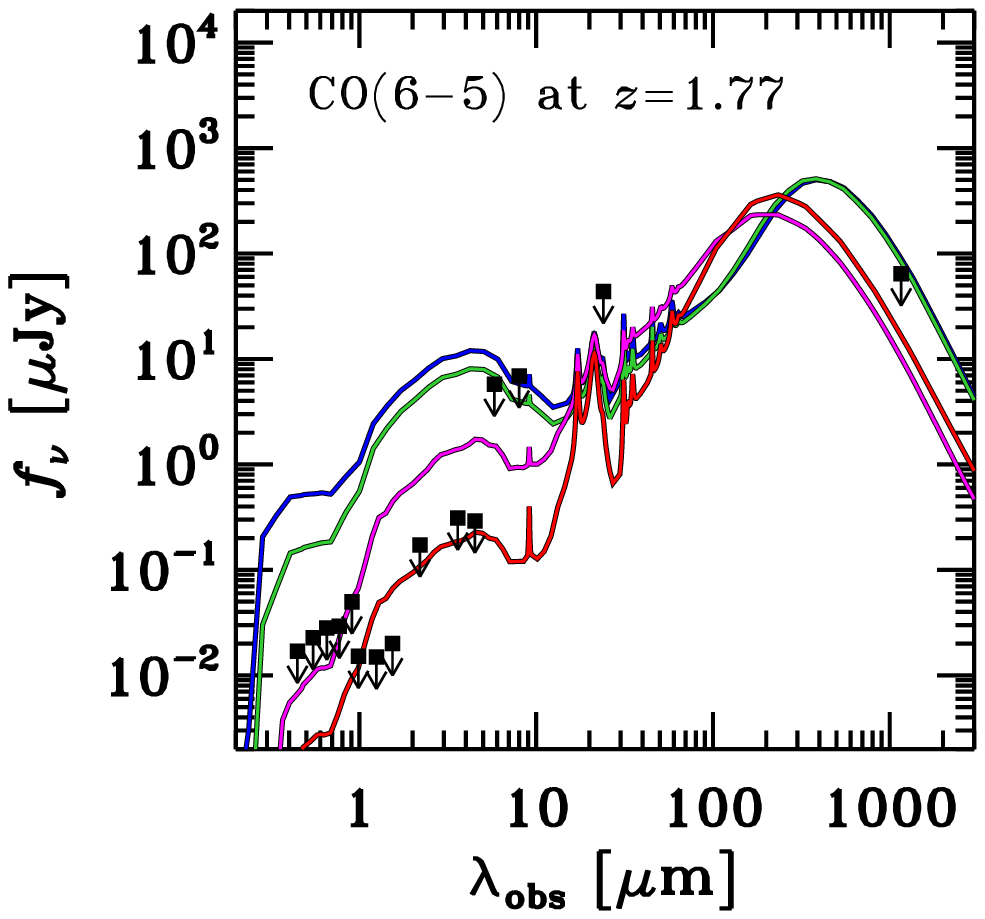}
   \includegraphics[scale=0.7]{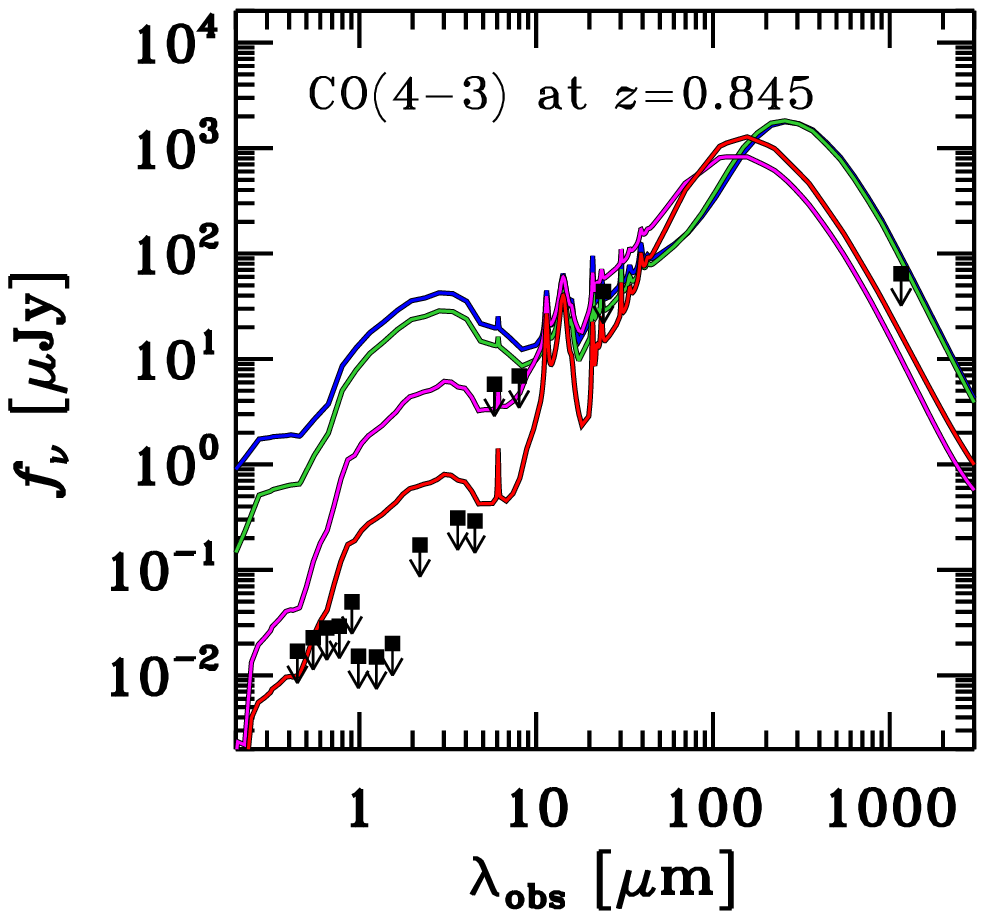}
 \caption[]
{
Multiwavelength SED of SLE--1. 
The vertical arrows with filled squares show upper limits for 
Subaru/Suprime-Cam $B$, $V$, $R$, $i'$, $z'$, 
\textit{HST}/WFC3 F098M, F125W, F160W, 
UKIRT/WFCAM $K$, 
\textit{Spitzer}/IRAC $3.6\mu$m, $4.5\mu$m, $5.8\mu$m, $8.0\mu$m, 
\textit{Spitzer}/MIPS $24\mu$m, 
and ALMA Band 6 
from left to right. 
For comparison, 
we plot 
the redshifted SEDs of Arp220 (red line), 
M82 (magenta line), 
M51 (green line), 
and NGC 6946 (blue line) from \cite{silva1998}, 
scaled to 
the expected FIR luminosities 
from the detected line flux 
for four cases: 
[{\sc Cii}]$158\mu$m (upper-left), 
CO(8--7) (upper-right), 
CO(6--5) (bottom-left), 
and CO(4--3) (bottom-right).
The upper limits estimated from the \textit{Herschel} SPIRE data are not shown, 
since they are shallow 
and affected by source confusion as shown in Figure \ref{fig:sme1_face_FIR}. 
}
\label{fig:sed_opt_mm}
\end{center}
\end{figure*}

Next, 
taking advantage of the upper limits estimated from the multiwavelength observations, 
we discuss what the detected line of SLE--1 is. 
Figure \ref{fig:sed_opt_mm} 
shows the multiwavelength SED of SLE--1, 
together with the SED templates of 
dusty starbursts (Arp220 and M82) 
and 
spiral galaxies (M51 and NGC 6946) 
taken from \cite{silva1998}. 
The SED templates are normalized to 
the expected FIR luminosities $L_{\rm IR}^{\rm (exp)}$ presented in Table \ref{tab:line_candidates}. 
As can be seen in Figure \ref{fig:sed_opt_mm}, 
the tight upper limits on the flux densities 
from the optical to the near-infrared wavelengths 
rule out the possibilities that the detected line is the CO lines.
If we adopt the CO excitation ladder of Milky Way \citep[Figure 4 of][]{carilli2013} 
instead of the M82 ladder, 
we obtain higher FIR luminosities 
and thus larger normalization factors of the expected galaxy SEDs 
than those with the M82 CO ladder, 
which are ruled out by the upper limits 
from the optical to near-infrared. 
The possibility of [{\sc Ci}](1--0) is also excluded 
by the upper limits at the optical and the near-infrared wavelengths. 
In the case of [{\sc Cii}]$158\mu$m, 
the SEDs of the blue galaxies (M51 and NGC 6946) are ruled out, 
while 
the SEDs of the galaxies with red colors (Arp220 and M82) 
are consistent with the upper limits 
on the flux densities at wavelengths from the optical to $1.2$ mm.

In summary, 
the possibilities of the CO lines and the [{\sc Ci}](1--0) line 
have been excluded,  
but that of the [{\sc Cii}] line with the red SEDs (Arp220 and M82) 
is not ruled out. 
Finally, 
we examine another possibility:  
the possibility that the detected line is [{\sc Oiii}]$88\mu$m 
from a galaxy at $z = 12.6$.  
Since [{\sc Oiii}]$88\mu$m is one of the strongest FIR lines 
from {\sc Hii} regions, 
\cite{inoue2014} have derived 
a line emissivity model for the [{\sc Oiii}]$88\mu$m line 
as a function of metallicity 
with the photoionization code {\sc cloudy} \citep{ferland2013}. 
We adopt the line emissivity model 
to estimate the total SFRs from the [{\sc Oiii}]$88\mu$m line luminosity,  
and 
test whether the estimated SFRs are consistent with 
the upper limits on the flux densities of SLE--1 
obtained from the $K$-band and ALMA Band 6 observations, 
which correspond to the rest-frame UV ($\sim 1600${\AA}) and FIR, respectively. 
We do not use the upper limit estimated from the WFC3 F160W image, 
since in this case, 
the redshifted Ly$\alpha$ and the continuum break 
enters the F160W band. 
The upper limit of the $K$-band data 
gives the upper limit of the unobscured SFR,  
SFR$_{\rm UV} \simeq 20 M_\odot$ yr$^{-1}$ \citep{kennicutt1998}.
Based on the ALMA Band 6 continuum data, 
the upper limit of the obscured SFR 
is estimated to be SFR$_{\rm IR} \simeq 40 M_\odot$ yr$^{-1}$ 
by the same method as noted in Section \ref{sec:introduction}. 
Thus, 
if the total SFR expected from the line flux 
is higher than their sum, 
SFR$_{\rm UV}$ + SFR$_{\rm IR}$ $\gtrsim 60 M_\odot$ yr$^{-1}$, 
the possibility of [{\sc Oiii}]$88\mu$m is excluded. 
Based on Figure 1 of \cite{inoue2014},  
if the galaxy metallicity is lower than $0.16 Z_\odot$ or higher than $0.32 Z_\odot$,  
the total SFRs estimated from the [{\sc Oiii}]$88\mu$m luminosity 
are higher than $60 M_\odot$ yr$^{-1}$. 
Thus, 
the possibility of [{\sc Oiii}]$88\mu$m is ruled out 
if the galaxy metallicity is $Z/Z_\odot \lesssim 0.2$ or $Z/Z_\odot \gtrsim 0.3$; 
[{\sc Oiii}]$88\mu$m is possible 
only in the cases of $Z/Z_\odot \simeq 0.2-0.3$.  
The estimated total SFR 
based on Figure 1 of \cite{inoue2014} 
has a minimum of about $50 M_\odot$ yr$^{-1}$ at $Z = 0.2 Z_\odot$. 
Note that many observational studies have reported that 
the number density of star-forming galaxies at high redshifts ($z \gtrsim 7$)
decreases with increasing redshift 
\citep[e.g.,][]{bouwens2011,ellis2012b,schenker2012,mclure2012b,coe2013,oesch2013b}. 
Although this implies that
it is unlikely 
that a rare high-redshift galaxy with an intense star formation of $50-60 M_\odot$ yr$^{-1}$ 
is identified in our small field of view,  
the possibility of [{\sc Oiii}]$88\mu$m cannot be excluded 
from the current observational data.

To summarize the above, 
we have 
reported 
the detection of SLE--1 
and 
discussed the possible interpretations for the line of SLE--1. 
We have found that 
the possible interpretations are 
[{\sc Cii}]$158\mu$m from a dusty star-forming galaxy at $z=6.60$  
or 
 [{\sc Oiii}]$88\mu$m from a moderately metal-enriched star-forming galaxy at $z=12.6$. 
If the detected line is [{\sc Cii}]$158\mu$m, 
SLE--1 would be at a similar redshift to Himiko. 
In this case, 
the galaxy would be at a projected distance of 
$\simeq 34$ proper kpc ($\simeq 260$ kpc in comoving units) from Himiko.  
These systems might merge into a single galaxy.

It should be noted that, however, that 
some recent studies of local LIRGs have revealed that 
shocks can produce exceptionally high CO line to FIR continuum luminosity ratios.  
For instance, 
\cite{meijerink2013} have reported that 
a nearby LIRG, NGC 6240, has a CO-to-continuum luminosity ratio 
about an order of magnitude higher than the typical ratio of local ULIRGs such as Mrk231 and Arp220.  
If shock excitation is exceptionally effective in SLE--1 
and the CO-to-continuum luminosity ratio is an order of magnitude higher than 
what is expected from the relationship adopted above, 
the expected FIR luminosities from the observed line flux 
are estimated to be substantially fainter. 
In this case, 
the possibilities of high-$J$ CO transition lines with red SEDs would not be 
ruled out.

Although 
SLE--1 shows the significant line detection, 
again we do not rule out the possibility that 
it is 
caused by unknown systematic noise effects as discussed above. 
Our carried-over ALMA cycle 1 program for Himiko  
will observe 
SLE--1. 
The carried-over observations will give us an opportunity 
to carefully test whether 
SLE--1 is 
real or not.

\section{Summary} \label{sec:summary}

In this paper, 
we have presented the number counts and the spatial clustering 
of faint SMGs,  
and 
reported 
a serendipitous detection of an SLE  
with no multiwavelength continuum counterpart 
revealed by the deep ALMA observations. 
Exploiting the deep ALMA Band 6/Band 7 continuum data 
for the $10$ independent fields that reduce the effect of cosmic variance, 
we have detected faint SMGs with flux densities of $0.1-1.0$ mJy. 
In addition, 
we have conducted a blind search for line emitters 
in the ALMA data cubes, and identified 
SLE--1. 
Our main results are as follows.

\begin{itemize}

\item We have constructed 
the $1.2$ mm differential number counts of SMGs 
and found that 
the number counts increase 
with decreasing flux density down to $0.1$ mJy. 
We have also found that  
the slope of the number counts for the faint 
($0.1-1$ mJy, or SFR$_{\rm IR} \sim 30-300 M_\odot$ yr$^{-1}$) SMGs 
is smaller than that for bright ($>1$ mJy) SMGs. 
Our number counts have revealed that 
the faint SMGs contribute about $50${\%} of the EBL, 
which is significantly larger than the contributions from the bright SMGs ($\sim 7${\%}). 
The remaining $40${\%} of the EBL is contributed by 
very faint SMGs with flux densities of $<0.1$ mJy 
(SFR$_{\rm IR} \lesssim 30 M_\odot$ yr$^{-1}$).

\item From the field-to-field scatter in their number counts, 
we have obtained a coarse estimate of the galaxy bias of the faint SMGs, $b_{\rm g} < 4$, 
which suggests that  
the dark halo masses of the faint SMGs is $M_{\rm DH} \lesssim 7 \times 10^{12} M_\odot$. 
Their bias 
is found to be lower than those of bright SMGs 
(\citealt{webb2003}; \citealt{blain2004}; \citealt{weiss2009}; \citealt{hickox2012}; c.f., \citealt{williams2011}), 
indicating the clustering segregation with the FIR luminosity in SMGs. 
We also find that 
the galaxy bias of the faint SMGs is consistent with those of 
abundant star-forming galaxy populations at high redshifts 
such as sBzKs, LBGs, and LAEs,  
which implies that 
some of the faint SMGs might be their FIR counterparts. 
It should be noted that 
our estimates suffer from relatively large uncertainties 
mainly due to the small number statistics 
and unexplored redshift distribution of the faint SMG population, 
which will be overcome 
after a large number of deep ALMA maps and 
a redshift distribution of faint SMGs become available 
in the near future.

\item We have found that 
SLE--1 has 
no counterpart in the multiwavelength images, 
suggesting that 
it would be a faint galaxy at a high redshift. 
SLE--1 
shows a significant line detection with an SNR of $7.1$ at $249.9$ GHz.  
Taking advantage of the upper limits estimated 
from the deep images at wavelengths from the optical to $1.2$ mm,  
we have discussed  
what the detected line and the redshift of SLE--1 are. 
If the detection of SLE--1 is not induced by unknown systematic noise effects in ALMA data, 
the possible explanations for the detected line of SLE--1 are 
[{\sc Cii}]$158\mu$m  from a dusty star-forming galaxy at $z=6.60$   
or 
[{\sc Oiii}]$88\mu$m from a star-forming galaxy 
with a moderate metallicity of $Z/Z_\odot \simeq 0.2-0.3$ at $z=12.6$. 

\end{itemize}

\section*{Acknowledgements}

We thank 
Bunyo Hatsukade 
for giving us helpful advice on analyzing the data and sending their results. 
We also thank 
Christopher C. Hayward, Ikko Shimizu, and Zhen-Yi Cai 
for giving us their model predictions of the differential number counts. 
We appreciate the support of the staff at the ALMA Regional Center, 
especially Kazuya Saigo, Daisuke Iono, and Shinya Komugi.  
We are grateful to 
Matthieu B{\'e}thermin, 
Chian-Chou Chen, 
Christina C. Williams, 
Bunyo Hatsukade, 
Zhen-Yi Cai 
and 
Seiji Fujimoto 
for their comments on earlier versions of the manuscript. 
We are grateful to the anonymous referee for valuable comments and suggestions 
which significantly improved the manuscript. 
This work was supported by 
Japan Society for the Promotion of Science (JSPS), 
KAKENHI, 
Grants-in-Aid for Research Activity Start-up Grant Number 24840010  
and 
Grant-in-Aid for Scientific Research (A) Grant Number 23244025. 
This work was also supported by 
World Premier International Research Center Initiative (WPI Initiative), MEXT, Japan. 
M.O. was supported by the ALMA Japan Research Grant
of NAOJ Chile Observatory, NAOJ-ALMA-0005.
This paper makes use of the following ALMA data: 
ADS/JAO.ALMA{\#}2011.0.00115.S,  
{\#}2012.1.00602.S,  
{\#}2011.0.00206.S,  
{\#}2011.0.00243.S,  
and
{\#}2011.0.00268.S. 
ALMA is a partnership of ESO (representing its member states), 
NSF (USA) and NINS (Japan), together with NRC (Canada) and NSC and ASIAA (Taiwan), 
in cooperation with the Republic of Chile. 
The Joint ALMA Observatory is operated by ESO, AUI/NRAO and NAOJ.


\bibliographystyle{apj}
\bibliography{apj-jour,../../../Papers/papers_gal/ms}


\end{document}